\documentclass{aa}
\usepackage{graphicx}
\usepackage{txfonts}
\usepackage{lipsum}
\usepackage{subcaption} 
\usepackage{lscape} 
\usepackage{placeins}         
\usepackage{xcolor}
\definecolor{cobalt}{rgb}{0.2118, 0.3176, 0.5804}
\usepackage{hyperref}    
\hypersetup{
  colorlinks=true,
  citecolor=cobalt,
  linkcolor=cobalt,
  urlcolor=cobalt,
}
\usepackage{float}
\usepackage{stfloats}
\usepackage{cleveref} 
\Crefname{equation}{Eq.}{Eqs.}
\Crefname{figure}{Fig.}{Figs.}
\usepackage{orcidlink}
\usepackage{siunitx}
\usepackage{multirow}
\usepackage{comment}
\usepackage{caption}
\def\dd{\mathrm{d}}
\def\l{\left}
\def\r{\right}
\def\f{\frac}
\def\chimera{\textsc{chimera}\xspace}
\def\icaro{\textsc{icarogw}\xspace}
\def\pymcpop{\textsc{pymcpop-gw}\xspace}

\newcommand{\condmid}{\mathrel{\mspace{0.05mu}|\mspace{0.05mu}}}

\DeclareSIUnit \parsec {pc}
\DeclareSIUnit \years{yrs}
\newcommand{\msun}{\ensuremath{\mathrm{M}_\odot}}

\begin{document}

\title{Pushing spectral siren cosmology into the third-generation era: \\ A blinded mock data challenge}

\author{Matteo Tagliazucchi\inst{1,2,3}
\corrauth{\href{emailto:matteo.tagliazucchi2@unibo.it}{matteo.tagliazucchi2@unibo.it}}, \orcidlink{0009-0003-8886-3184}, 
Michele Moresco\inst{1,3}, \orcidlink{0000-0002-7616-7136}, 
Alessandro Agapito\inst{4}, \orcidlink{0009-0005-9004-3163}, 
Michele Mancarella\inst{4}, \orcidlink{0000-0002-0675-508X},\\ 
Sarah Ferraiuolo\inst{5,6,7}, \orcidlink{0009-0005-5582-2989}, 
Simone Mastrogiovanni\inst{5}, \orcidlink{0000-0003-1606-4183}, 
Nicola Borghi\inst{1,2,3}, \orcidlink{0000-0002-2889-8997}, \\
Francesco Pannarale\inst{5,6}, \orcidlink{0000-0002-7537-3210}, and
Daniele Bonacorsi\inst{1,2}\orcidlink{0000-0002-0835-9574}
} 

\institute{Dipartimento di Fisica e Astronomia ``Augusto Righi''--Universit\`{a} di Bologna, Viale Berti Pichat 6/2, I-40127 Bologna, Italy
\and
INFN - Sezione di Bologna, Viale Berti Pichat 6/2, I-40127 Bologna, Italy
\and
INAF - Osservatorio di Astrofisica e Scienza dello Spazio di Bologna, via Piero Gobetti 93/3, I-40129 Bologna, Italy
\and 
Aix-Marseille Universit\'e, Universit\'e de Toulon, CNRS, CPT, Marseille, France
\and 
INFN, Sezione di Roma, Piazzale Aldo Moro 5, I-00185 Roma, Italy
\and
``Sapienza'' Universit\'a di Roma, Piazzale Aldo Moro 5, 00185 Roma, Italy
\and
Aix Marseille Univ, CNRS/IN2P3, CPPM, Marseille, France
}

\date{Received 19 February 2026 / Accepted 19 May 2026} 

\abstract{
Gravitational wave (GW) spectral sirens offer a promising method for measuring cosmological parameters using GW data only --- without relying on external redshift information such as electromagnetic counterparts or galaxy catalogs --- by exploiting distributional features in the population of GW sources. 
The advent of third-generation detectors such as the Einstein Telescope (ET) will provide catalogs three orders of magnitude larger than current ones, raising questions about the scalability and robustness of existing inference pipelines. 
We present a blinded mock data challenge that tests three public pipelines with distinct numerical implementations, namely \icaro, \chimera, and \pymcpop, on simulated ET observations containing the best $\mathcal{O}(10^4)$ binary black hole mergers that can be observed in 1 year. 
We assessed their computational performance, validated their agreement in a blinded setting, and forecast cosmological constraints. 
We find that, thanks to GPU acceleration, these pipelines can process the events expected from ET within a manageable timeframe. 
All pipelines recover consistent cosmological and population parameters. 
Assuming a flat Lambda cold dark matter ($\Lambda$CDM) model, we measure $H(z)$ at $z\sim1.5$ with 2.4\% precision, and achieve a mean precision on $H(z)$ of 2.8\% across $0.7<z< 1.8$ with a catalog of $\sim 12,000$ high S/N events. 
This corresponds to joint constraints of $\sim 10\%$ on $H_0$ and $\sim 26\%$ on $\Omega_{\mathrm{m},0}$.
We also identified the events that contribute mostly to constraining cosmological parameters, and show that low-distance sources near population features drive the constraining power on all cosmological parameters, while higher-distance events primarily constrain $\Omega_{\rm m,0}$. 
Our results establish a validated, performance-tested framework for spectral siren cosmology in the era of third-generation GW observatories.
}


\titlerunning{Pushing spectral siren cosmology into the third-generation era: A blinded mock data challenge}

\authorrunning{Tagliazucchi et al}

\maketitle
\nolinenumbers

\section{Introduction}
Since the first direct detection of gravitational waves (GWs) by the LIGO-Virgo Collaboration in 2015 \citep{LIGOScientific:2016aoc}, GW observations have opened a new observational window, providing cosmological information complementary to that of electromagnetic (EM) probes. GWs can be used as standard cosmological rulers as they allow   a direct measurement of the luminosity distance to the source, independent of the traditional cosmic distance ladder. 
When combined with redshift information, they enable constraints on cosmological parameters, for example the Hubble constant \citep{Schutz:1986gp, Holz:2005df, Moresco:2022phi, Jin:2025dvf}, thereby probing the expansion history of the Universe across cosmic time.

A fundamental challenge for GW-based cosmology is the degeneracy between the source redshift and the intrinsic mass scale of the binary system. 
Several methods have therefore been developed to infer redshift information statistically. 
One prominent approach is the dark-siren method, which relies on cross-correlating the GW sky localization volume with galaxy catalogs and assigning probabilistic weights to potential host galaxies \citep{DelPozzo:2011vcw, Chen:2017rfc, LIGOScientific:2018gmd, Gray:2019ksv, Finke:2021aom, Gray:2021sew, Belfiglio:2022cnd, Gair:2022zsa}. 

For this work we focused on the spectral siren method, which exploits features in the population distribution of GW source parameters, most notably the mass distribution, to statistically disentangle redshift and intrinsic properties across the detected population \citep{Taylor:2011fs, Farr:2019rap, Mastrogiovanni:2021wsd}.
This approach allows the joint inference of cosmological and population hyperparameters without relying on external redshift measurements.

Several analysis pipelines have been developed to implement this joint inference framework, including \icaro \citep{Mastrogiovanni:2023emh, Mastrogiovanni:2023zbw}, \textsc{gwcosmo} \citep{Gray:2023wgj}, \chimera \citep{Borghi:2023opd, Tagliazucchi:2025ofb}, and \pymcpop \citep{Mancarella:2025uat}. 
Dark and spectral siren techniques have already been applied to current LIGO–Virgo–KAGRA observations, and have yielded constraints on the Hubble constant \citep{LIGOScientific:2021aug, LIGOScientific:2019zcs,LIGOScientific:2025jau} from the publicly available GWTC-4.0 catalog and its previous versions \citep{LIGOScientific:2018mvr, LIGOScientific:2020ibl, KAGRA:2021vkt, LIGOScientific:2025slb}. 
These measurements, however, remain limited by detector sensitivity and the relatively small number of detected events.

This situation will change substantially with the advent of third-generation (3G) GW detectors. 
Two concepts are currently under consideration for construction in the 2030s: the Einstein Telescope (ET) in Europe \citep{Hild:2010id, Punturo:2010zz, ET:2019dnz} and the Cosmic Explorer (CE) in the United States \citep{Reitze:2019iox, Evans:2021gyd}. 
Their improved sensitivity and extended low-frequency coverage will enable the detection of compact binary mergers over a broad redshift range, potentially reaching the dark ages, with expected detection rates of up to millions of events over a timescale of approximately ten years \citep{ET:2019dnz, Iacovelli:2022bbs, Branchesi:2023mws,ET:2025xjr}

In this high-statistics regime, the accuracy and precision of cosmological constraints will be driven not only by detector sensitivity, but also by the behavior of inference pipelines when confronted with large and heterogeneous datasets.
Key questions include how constraints scale with the number of detected events, and whether analysis pipelines remain stable and unbiased as the dimensionality and volume of the data increase. 
These issues involve convergence properties, sampling efficiency, and robustness to modeling assumptions.
Moreover, a critical question is whether current cosmological pipelines can scale to accommodate the expected future number of detected GW events, which are projected to increase by three orders of magnitude.
Notably, different spectral siren pipelines implement the hierarchical likelihood using distinct numerical strategies. 
Therefore, ensuring that these different numerical implementations produce consistent and unbiased results is a key aspect of validation for the 3G era.

To address these challenges, we performed a blinded mock data challenge (MDC) designed to test some of the publicly available spectral siren pipelines under realistic observational conditions expected for ET. 
The use of blinded analyses allows an unbiased validation of inference pipelines and enables a controlled comparison of different implementations. 
This study represents a first stress test of these codes at the data volume expected from third-generation detectors, providing both a performance comparison and a forecast of the execution times required for future analyses.
Moreover, this study features the first direct comparison between codes that compute the likelihood for inferring cosmological parameters from GW data in three different ways, allowing us to validate their numerical consistency and to identify potential systematic differences.
Additionally, we investigate the contribution of GW events to the reconstruction of the cosmic expansion history, thereby identifying the regimes in the observed parameter space in which the constraining power of spectral sirens is strongest.

This paper is organized as follows. 
In \Cref{sec:Framework} we introduce the statistical framework for spectral siren inference and describe the analysis pipelines employed in this study, including implementations based on \icaro, \chimera, and \pymcpop. 
In \Cref{sec:Blinded mock catalogs} we detail the construction of the blinded mock catalogs and the design of the mock data challenge. 
In Section \ref{sec:Results} we present the results of the MDC, focusing on pipeline performance, convergence behavior, and cosmological constraints. 
We conclude in \Cref{sec:Conclusion} with a discussion of the implications of our findings and possible future developments.

\section{Framework}
\label{sec:Framework}
Constraining cosmological and population parameters $\boldsymbol{\Lambda}$ (such as $H_0$, merger rate density, and mass distribution) from a set of GW observations $\{\boldsymbol{d}_i\}_{i=1}^{N_{\rm obs}}$ is a hierarchical problem, as it requires inferring population-level parameters from the measured detector-frame properties of individual events $\{\boldsymbol{\theta}_{\dd,i}\}_{i=1}^{N_{\rm obs}}$ (e.g., redshifted masses, and luminosity distance).
The hierarchical likelihood of obtaining $N_{\rm obs}$ observations over an observing time $T_{\rm obs}$ in the presence of a selection effect is \citep[see, e.g.,][for a detailed derivation]{Mandel:2018mve, Vitale:2020aaz}
\begin{align}
    \mathcal{L}\l(\{\boldsymbol{d}_i\}_{i=1}^{N_{\rm obs}} \condmid \boldsymbol{\Lambda}\r) \propto e^{-N_{\rm exp}(\boldsymbol{\Lambda})} \times \nonumber \\ 
    \prod_{i=1}^{N_{\rm obs}} T_{\rm obs} \!\int\! \dd \boldsymbol{\theta}_{\dd,i}  \mathcal{L}_{\rm gw}\l( \boldsymbol{d}_i  \condmid  \boldsymbol{\theta}_{\dd,i}\r)  \f{\dd t_i}{\dd t_{\rm d, i}} \left|\f{\dd \boldsymbol{\theta}_i}{\dd \boldsymbol{\theta}_{\dd,i}} \right| \f{\dd N}{\dd t_i\dd \boldsymbol{\theta}_i}(\boldsymbol{\Lambda}) \equiv \nonumber \\ 
    \equiv e^{-N_{\rm exp}(\boldsymbol{\Lambda})} \prod_{i=1}^{N_{\rm obs}} T_{\rm obs} I_i \label{eq:hierarchical-lkl} \;,
\end{align}
where $\mathcal{L}_{\rm gw}$ is the single GW event likelihood, $\f{\dd N}{\dd t \dd \boldsymbol{\theta}}(\boldsymbol{\Lambda})$ is the population model describing the rate of events with source-frame parameters $\boldsymbol{\theta}$, and $|\f{\dd \boldsymbol{\theta}}{\dd \boldsymbol{\theta}_\dd}|$ is the Jacobian computed from the source-frame to the detector-frame.
The parameter $t$ is the coalescing time in source-frame and $t_{\dd}$ the corresponding detector-frame time, and the factors $\f{\dd t}{\dd t_\dd}$ accounts for their conversion.
The $N_{\rm exp}(\boldsymbol{\Lambda})$ term represents the expected number of detectable sources over $T_{\rm obs}$, given the population model:
\begin{equation}\label{eq:n-exp}
    N_{\rm exp}(\boldsymbol{\Lambda}) = T_{\rm obs} \!\int\! \dd \boldsymbol{\theta}_\dd P_{\rm det}(\boldsymbol{\theta}_\dd) \f{\dd t}{\dd t_d}\l|\f{\dd \boldsymbol{\theta}}{\dd \boldsymbol{\theta}_\dd} \r|\f{\dd N}{\dd t \dd \boldsymbol{\theta}}(\boldsymbol{\Lambda})\; .
\end{equation}
Here $P_{\rm det}(\boldsymbol{\theta}_\dd)$ is the probability that an event with parameters $\boldsymbol{\theta}_\dd$ is detected by the considered detector network.  

In the spectral siren case, {dropping the assumption that GW events of interest here originate in galaxies, the population model and the Jacobian are given by
\begin{align}\label{eq:spectral-sirens-rate}
    \f{\dd N}{\dd t\dd \boldsymbol{\theta}}(\boldsymbol{\Lambda}) &= R_0 \psi(z\condmid \boldsymbol{\lambda}_r) \f{\dd V_c}{\dd z}(z\condmid \boldsymbol{\lambda}_c) p(m_1, m_2 \condmid \boldsymbol{\lambda}_m)\; , \\
     \l|\f{\dd \boldsymbol{\theta}}{\dd \boldsymbol{\theta}_\dd}\r| &= \l|\f{\partial d_L(z\condmid \boldsymbol{\lambda}_c)}{\partial z} \r|^{-1}(1+z)^{-2}\; , \\
     \f{\dd t}{\dd t_\dd} &= (1+z)^{-1} \; ,
\end{align}
where $R_0$ is the source rate per unit comoving volume per year, $\psi(z\condmid \boldsymbol{\lambda}_r)$ describes the merger rate evolution, parameterized by $\lambda_r$; $\dd V_c / \dd z$ is the comoving volume element, which depends on cosmological parameters $\lambda_c$; and $p(m_1, m_2 \condmid \boldsymbol{\lambda}_m)$ is the mass distribution with hyperparameters $\lambda_m$ \citep[see][for more details]{Mastrogiovanni:2023zbw}.

The hierarchal formalism described by \Cref{eq:hierarchical-lkl} depends on the detector-frame population rate, $\f{\dd N}{\dd t_\dd\dd \boldsymbol{\theta}_\dd}$.  
The population and cosmological parameters are inferred by (i) assuming a source-frame population rate $\f{\dd N}{\dd t\dd\boldsymbol{\theta}}$, (ii) propagating it to the detector frame via a cosmological model, and (iii) comparing the resulting detector-frame rate against the observed data \citep{Pierra:2025hoc}. 
Features in the population model, such as peaks, gaps, or breaks in the mass distribution, act as cosmological rulers and add information to constrain the cosmological parameters.
These features, which statistically break the mass-redshift degeneracy, are those that mostly drive the constraining power in the spectral siren approach.

We now describe three codes that compute the likelihood in \Cref{eq:hierarchical-lkl} differently.
The first code, \icaro, calculates the likelihood using a Monte Carlo sum, as done in \texttt{MGCosmoPop} \citep{Mancarella:2021ecn}.
Instead, \chimera uses a kernel density estimate (KDE) approach to handle the integrals in \Cref{eq:hierarchical-lkl}, analogously to \textsc{gwcosmo} \citep{Gray:2023wgj}.
Finally, \pymcpop implements a version of \Cref{eq:hierarchical-lkl} that is not marginalized over the event parameters.

\subsection{\icaro implementation}
In the \icaro pipeline, the integrals $I_i$ in \Cref{eq:hierarchical-lkl} are approximated using Monte Carlo integration
\begin{equation}\label{eq:Ii-icaro}
    I_i = \f{1}{N_{\rm pe}} \sum_{k=1}^{N_{\rm pe}} \f{1}{\pi(\boldsymbol{\theta}^k_{\dd,i})} \f{\dd t^k_i}{\dd t^k_{\dd, i}}\l|\f{\dd \boldsymbol{\theta}^k_i}{\dd \boldsymbol{\theta}^k_{\dd,i}}\r|\f{\dd N}{\dd t^k_i\dd \boldsymbol{\theta}^k_i}(\boldsymbol{\Lambda}) \equiv \f{1}{N_{\rm pe}} \sum_{k=1}^{N_{\rm pe}} n^k_i \; , 
\end{equation}
where we used the Bayes theorem $\mathcal{L}_{\rm gw}\l( \boldsymbol{d}_i \condmid  \boldsymbol{\theta}_{\dd,i}\r) = p_{\rm gw}\l( \boldsymbol{\theta}_{\dd,i} \condmid \boldsymbol{d}_i\r) / \pi(\boldsymbol{\theta}_{\dd,i})$, and $\boldsymbol{\theta}^k_{\dd,i}$ are posterior estimate (PE) samples of the $i$-th event drawn from $p_{\rm gw}$.
The expected number of sources $N_{\rm exp}$ is estimated via an injection campaign: $N_{\rm inj}$ events are drawn from a prior $\pi_{\rm inj}$ and processed through a detection pipeline. 
The number of detected events $N_{\rm det}$ is used to approximate \Cref{eq:n-exp} as a Monte Carlo sum over detected events \citep{Tiwari:2017ndi,LIGOScientific:2020kqk}:
\begin{equation}\label{eq:n-exp-approx}
     N_{\rm exp}(\boldsymbol{\Lambda}) = \f{1}{N_{\rm inj}} \sum_{j=1}^{N_{\rm det}} \f{1}{\pi_{\rm inj}(\boldsymbol{\theta}_{\dd,j})} \f{\dd t_j}{\dd t_{\dd,j}} \l|\f{\dd \boldsymbol{\theta}_j}{\dd \boldsymbol{\theta}_{\dd,j}}\r|\f{\dd N}{\dd t_j\dd \boldsymbol{\theta}_j}(\boldsymbol{\Lambda})\; .
\end{equation}
For a detailed explanation of its functionality, we refer to \cite{Mastrogiovanni:2023zbw}.

\subsection{\chimera implementation}
In the \chimera code, \Cref{eq:n-exp} is approximated as in the \icaro pipeline (see \Cref{eq:n-exp-approx}), whereas the integrals $I_i$ appearing in \Cref{eq:hierarchical-lkl} are evaluated differently.
Using Bayes' theorem and the detector-to-source-frame transformation $\dd \boldsymbol{\theta}_{\dd,i} p_{\rm gw}(\boldsymbol{\theta}_{\dd,i} \condmid \boldsymbol{d}_i) = \dd \boldsymbol{\theta}_i p_{\rm gw}(\boldsymbol{\theta}_i(\boldsymbol{\theta}_{\dd,i}, \boldsymbol{\lambda}_c) \condmid \boldsymbol{d}_i)$, the integrals $I_i$ become
\begin{align}
    I_i = R_0 \!\int\! \dd \boldsymbol{\theta}_i \f{p_{\rm gw}(\boldsymbol{\theta}_i(\boldsymbol{\theta}_{\dd,i}, \boldsymbol{\lambda}_c) \condmid \boldsymbol{d}_i)}{\pi_{\rm gw}(\boldsymbol{\theta}_{\dd,i})} \nonumber \times \\ p(z_i, \hat\Omega_i \condmid \boldsymbol{\Lambda}) p(\boldsymbol{\tilde \theta}_i \condmid z_i, \hat\Omega_i, \boldsymbol{\Lambda}) \f{\dd t_i}{\dd t_{\dd,i}} \l|\f{\dd \boldsymbol{\theta}_i}{\dd \boldsymbol{\theta}_{\dd,i}} \r| \nonumber = \\
    = R_0 \!\int\! \dd z_i \dd \hat\Omega_i \mathcal{K}_{\rm gw}(z_i,\hat\Omega_i  \condmid \boldsymbol{d}_i, \boldsymbol{\Lambda}) p(z_i, \hat\Omega_i \condmid \boldsymbol{\Lambda})\; .
\end{align}
Here the source-frame population rate has been factorized into the overall normalization $R_0$, a term depending on redshift $z$ and sky-localization area $\hat\Omega$, and a term involving the remaining source parameters $\boldsymbol{\tilde \theta}$.
In the spectral siren case, the $\boldsymbol{\tilde \theta}$ parameters are only binary masses (see \Cref{eq:spectral-sirens-rate}).
The GW kernel $\mathcal{K}_{\rm gw}(z_i,\hat\Omega_i  \condmid \boldsymbol{d}_i, \boldsymbol{\Lambda})$ is evaluated using a weighted KDE:
\begin{align}
    \mathcal{K}_{\rm gw}(z_i,\hat\Omega_i \condmid \boldsymbol{d}_i, \boldsymbol{\Lambda})  = \!\int\! \dd \boldsymbol{\tilde \theta}_i \f{p_{\rm gw}(\boldsymbol{\theta}_i(\boldsymbol{\theta}_{\dd,i}, \boldsymbol{\lambda}_c) \condmid \boldsymbol{d}_i)}{\pi_{\rm gw}(\boldsymbol{\theta}_{\dd,i})}  \nonumber \times \\ p(\boldsymbol{\tilde \theta}_i \condmid z_i, \hat\Omega_i, \boldsymbol{\Lambda}) \f{\dd t_i}{\dd t_{\dd,i}} \l|\f{\dd \boldsymbol{\theta}_i}{\dd \boldsymbol{\theta}_{\dd,i}} \r| = \nonumber \\
    = \mathrm{KDE}\l[ (z^k_i, \hat\Omega^k_i) \condmid w^k_i \r]\; . \label{eq:chimera-kde}
\end{align}
The KDE is built on the PE samples $(z^k_i, \hat\Omega^k_i)$ drawn from the marginalized GW distribution of the $i$-th event $p_{\rm gw}(z_i,\hat\Omega_i \condmid \boldsymbol{\tilde \theta}_i, \boldsymbol{d}_i, \boldsymbol{\lambda}_c)$, each weighted by
\begin{equation}\label{eq:weights-kde-chimera}
    w^k_i = \f{p(\boldsymbol{\tilde \theta}^k_i \condmid z^k_i, \hat\Omega^k_i, \boldsymbol{\Lambda})}{\pi_{\rm gw}(\boldsymbol{\theta}^k_{\dd,i})} \f{\dd t^k_i}{\dd t^k_{\dd,i}} \l|\f{\dd \boldsymbol{\theta}^k_i}{\dd \boldsymbol{\theta}^k_{\dd,i}} \r|\; .
\end{equation} 
The KDE is then evaluated on the integration volume of the $i$-th event.
In \chimera there are three different ways to efficiently build and evaluate the KDE in \Cref{eq:chimera-kde}.
For the spectral siren analyses presented in this work, the single-1d binned method  was used (see \citealt{Tagliazucchi:2025ofb} for the detailed description of the algorithm).

\subsection{\pymcpop implementation}

The package \pymcpop follows the strategy introduced by \citet{Mancarella:2025uat}, which avoids the Monte Carlo marginalization over single-event parameters that appears in the standard implementation of \Cref{eq:hierarchical-lkl}.
Instead of integrating over $\boldsymbol{\theta}_{\dd,i}$ at each likelihood call, \pymcpop samples the full hierarchical posterior in an enlarged parameter space, where the event-level parameters $\{\boldsymbol{\theta}_{\dd,i}\}$ are treated as latent variables alongside the hyper-parameters $\boldsymbol{\Lambda}$.

The key step is considering the full hierarchical posterior on $\{\boldsymbol{\Lambda}, \boldsymbol{\theta}_{\dd,i}\}_{i=1}^{N_{\rm obs}}$, or equivalently the likelihood
\begin{align}
\mathcal{L}\!\left(\{\boldsymbol{d}_i\}_{i=1}^{N_{\rm obs}} \condmid \boldsymbol{\Lambda}, \{\boldsymbol{\theta}_{\dd,i}\}_{i=1}^{N_{\rm obs}}\right)
\propto e^{-N_{\rm exp}(\boldsymbol{\Lambda})} \times \nonumber \\
\prod_{i=1}^{N_{\rm obs}} T_{\rm obs}
\frac{p_{\rm gw}(\boldsymbol{\theta}_{\dd,i} \condmid \boldsymbol{d}_i)}{\pi_{\rm gw}(\boldsymbol{\theta}_{\dd,i})}\,
\f{\dd t_i}{\dd t_{\dd,i}}\l|\frac{\dd \boldsymbol{\theta}_i}{\dd \boldsymbol{\theta}_{\dd,i}}\r|\,
\frac{\dd N}{\dd t_i\dd \boldsymbol{\theta}_i}(\boldsymbol{\Lambda})\; , \label{eq:pymcpop-joint}
\end{align}
and implementing $p_{\rm gw}(\boldsymbol{\theta}_{\dd,i}\condmid \boldsymbol{d}_i)$ as an effective (data-dependent) proposal and/or prior for each event, while the population information enters only through the reweighting factor multiplying it in \Cref{eq:pymcpop-joint}.
Compared to \icaro and \chimera, this eliminates the per-event Monte Carlo sum over PE samples in the numerator of \Cref{eq:hierarchical-lkl}, leaving selection effects as the only Monte Carlo ingredient through $N_{\rm exp}(\boldsymbol{\Lambda})$ (estimated from injections as in \Cref{eq:n-exp-approx}), at the cost of sampling a higher-dimensional posterior. 
This is achieved using a Hamiltonian Monte Carlo algorithm.

Since gradient-based samplers require continuous and differentiable densities, \pymcpop replaces the discrete PE samples with a continuous approximation of $p_{\rm gw}$.
Following \citet{Mancarella:2025uat}, the interpolation is performed in a reparameterized detector-frame space (using the detector frame chirp mass, $\mathcal{M}_c^\dd$; the mass ratio,  $q$;  the luminosity distance,  $d_L$; and log unbounded transforms) to avoid sharp boundaries.
The interpolator is built via a Gaussian mixture model, with the number of components selected by minimizing the Bayesian information criterion.

\section{Blinded mock catalogs}
\label{sec:Blinded mock catalogs}

To generate the mock catalogs used in this analysis, we model the population of binary black hole (BBH) sources as follows.
The merger rate distribution is assumed to follow the star formation history, described by the Madau–Dickinson formula \citep{Madau:2014bja}.
The mass distribution is a power law plus peak (PLP) law \citep[see Sec. C of][for more details]{LIGOScientific:2025jau}.
The underlying cosmological model is assumed to be flat Lambda cold dark matter ($\Lambda$CDM).
The summary of all hyperparameters considered is given in \Cref{tab:hyperparams_summary}.

The fiducial parameters for this population model are drawn from the constraints for this model from the GWTC-3 catalog \citep{LIGOScientific:2021aug}.
The posterior distribution from which the fiducial parameters were drawn is obtained with intentionally narrower priors on the cosmological parameters (specifically, $H_0\in[60,80]\,\mathrm{km/s/Mpc}$ and $\Omega_{\mathrm{m},0}\in[0.2,0.4]$) to ensure that the selected fiducial values remain physically plausible and avoid unrealistic extremes.
A strict blinding procedure is then implemented. The drawn fiducial parameters, along with the computed total number of coalescences in one year given the fiducial model, are immediately sent via an automated Python script to an external collaborator for safekeeping and are subsequently deleted from the origin server to ensure the blinding in the subsequent analysis.

\begin{figure*}[!htb]
    \centering
    \includegraphics[width=0.95\textwidth]{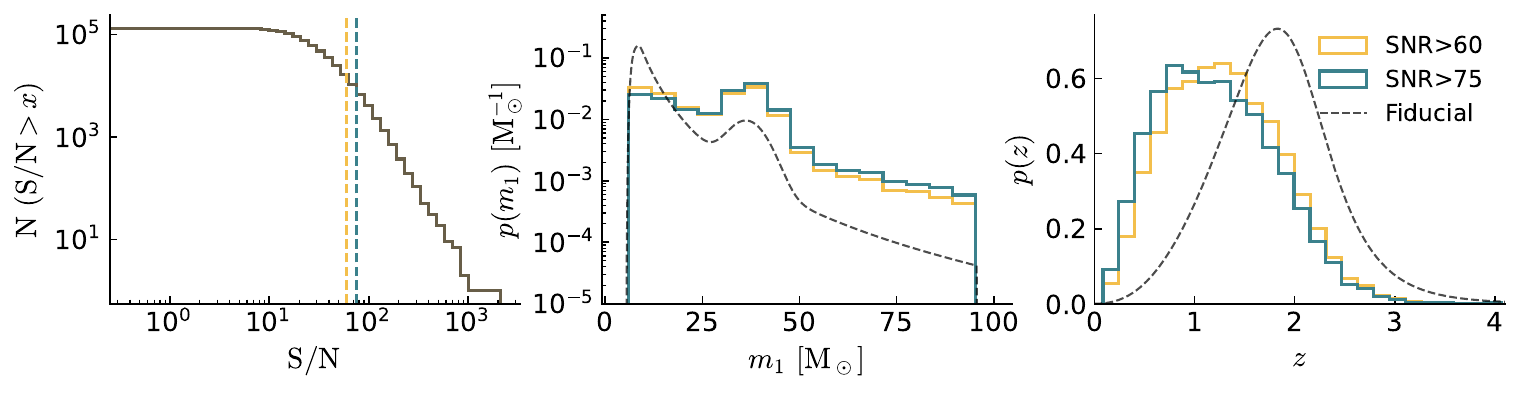}
    \caption{Properties of the blinded mock BBH catalogs. Left panel: Reverse cumulative distribution of the S/N for the generated population. The dashed lines indicate the S/N thresholds used to define the two catalogs. Middle and right panels: Mass and redshift rate \Cref{eq:p-z} distributions, respectively, for events in the two catalogs, compared with the fiducial population model (dashed lines).
    }
    \label{fig:cat-props}
\end{figure*}

Using the blinded fiducial model, we first compute the total number of binary coalescences expected per detector-frame year, $\f{\dd N_{\rm cbc}}{\dd t_\dd}$.
This is obtained by integrating the source-frame population rate \Cref{eq:spectral-sirens-rate} over all parameters except the coalescing time $t$.
Then, given the drawn fiducial model, the total number of sources is
\begin{align}\label{eq:event-rate}
    \f{\dd N_{\rm cbc}}{\dd t_\dd} & = \!\int\! \dd \boldsymbol{\theta} \f{\dd N}{\dd t_d \dd \boldsymbol{\theta}} =   \!\int\! \dd \boldsymbol{\theta} \f{\dd N}{ \dd t \dd \boldsymbol{\theta}} \f{\dd t}{\dd t_\dd} \nonumber \\
    & =   R_0 \!\int_0^{z_{\rm max}} \dd z \f{\psi(z ; \boldsymbol{\lambda}_r)}{1+z} \f{\dd V_c}{\dd z}(z; \boldsymbol{\lambda}_c) \approx 1.3\times 10^5,
\end{align}
where $z_{\rm max}$ is conservatively set to 100 to cover the whole detector horizon and ensure that $\psi(x;\boldsymbol{\lambda}_r) \approx 0$.
In the second line of the previous equation we used the normalization of the mass distribution, and the redshift-dependent factor $\dd t_\dd / \dd t = 1+z$ arising from cosmic expansion.
We then randomly draw the intrinsic parameters (masses, redshifts) for each of these $1.3\times 10^5$ sources from the fiducial population distributions. 
In particular, the redshift distribution, $p(z)$, describes the rate of event per unit of redshift and detector-frame time.
It depends on both cosmological and merger rate parameters, and it is calculated similarly to equation \Cref{eq:event-rate},
\begin{equation}\label{eq:p-z}
    \f{\dd N_{\rm cbc}}{\dd z \dd t_\dd} = \!\int\! \dd \boldsymbol{\overline{\theta}} \f{\dd N}{\dd \boldsymbol{\overline{\theta}} \dd t} \f{\dd t}{\dd t_\dd} = R_0 \f{\psi(z ; \boldsymbol{\lambda}_r)}{1+z} \f{\dd V_c}{\dd z}(z; \boldsymbol{\lambda}_c) \propto p(z) ,
\end{equation}
where $\boldsymbol{\overline{\theta}}$ are all event source-frame parameters except $t$ and $z$.
The remaining parameters describing a GW event (i.e., spin components and extrinsic parameters) are drawn from uniform distributions: the spin components are uniform in $[-1,1]$; the inclination angle is uniform in $\cos(\iota)$ over $[0,\pi]$; the polarization angle and coalescence phase are uniformly distributed in $[0,\pi]$ and $[0,2\pi]$, respectively; the coalescence time, expressed in units of fraction of a day, is uniform in $[0,1]$; and the sky position ($\alpha, \delta$) is drawn isotropically, with $\alpha$ uniform in $[0, 2\pi]$ and $\delta$ uniform in $\cos(\delta)$ over $[0, \pi]$.

To determine which of these sources would be detected, we approximate their waveform using the \texttt{IMRPhenomHM} \citep{London:2017bcn} model and we inject it into a noise realization of the detector network. 
For the latter, we assume a configuration of the ET made of two L-shaped $15\,\mathrm{km}$ long interferometers, and we consider its publicly available nominal sensitivity \citep{Hild:2010id}, which is the most updated one adopted in \cite{ET:2019dnz, ET:2025xjr}.\footnote{The sensitivity curve is available at \url{https://www.et-gw.eu/index.php/etsensitivities}}
One of the two detectors is assumed to be located   in Sardinia and the other in the Euregio Meuse-Rhine region;  they are rotated by $45^\circ$ with respect to each other.
We then compute the signal-to-noise ratio (S/N) of each source using \texttt{GWFAST} \citep{Iacovelli:2022bbs, Iacovelli:2022mbg}.

The S/N distribution of the drawn BBHs is shown in the left panel of \Cref{fig:cat-props}. 
Due to the high sensitivity of ET, the number of BBHs detected (S/N > 8) is $96\%$ of the total BBH population.
However, we selected two sub-catalogs, one consisting of 11896 events at S/N $>$ 60 and one with 6843 events at S/N $>$ 75 to ease the computing load of the analysis.
This dataset   then allows  the code performances to be extrapolated at the full expected data volume.
The primary mass and redshift distributions of these two sub-catalogs are given in the central and right panel of \Cref{fig:cat-props}.

For parameter estimation, we approximated the GW likelihood for each detected event using a Fisher information matrix (FIM) approach. 
This method models the posterior distribution as a multivariate Gaussian, the covariance matrix of which is the inverse of the FIM, computed with \texttt{GWFAST}. 
This approximation has been verified in previous analyses \citep[see e.g][]{Borghi:2023opd, Tagliazucchi:2025ofb}, and works for the very high S/N events considered in this study.
In particular, we used \texttt{emcee} \citep{Foreman-Mackey:2012any} to draw 5000 PE samples for each event from the multivariate Gaussian mentioned above.
During the Markov chain Monte Carlo (MCMC) sampling we imposed physical priors on GW parameters.
The total prior used in each MCMC includes uniform distributions in $[0,10^5]\,\msun$ and $[0,1/4]$ for the chirp mass $\mathcal{M}_c$ and the symmetric mass ratio $\eta$, respectively, while all other waveform parameters were bounded in the same physical ranges from which they were drawn.
The prior on the luminosity distance is proportional to $d^2_L$ and bounded in $[0,10^5]\,\si{\giga\parsec}$.
The prior also includes the Jacobian of the transformation $(\mathcal{M}_c, \eta) \to (m^\dd_1, m^\dd_2)$ to ensure that the binary masses are uniformly distributed.
We chose 5000 PE samples because this number is sufficient to satisfy the numerical stability criterion for the numerical integration of the integrals $I_i$ in \Cref{eq:hierarchical-lkl}.
In particular, for \icaro we checked at each MCMC step that the number of effective samples of each event is
\begin{equation}
    N^{\rm eff}_{\rm PE, i} = \f{\l( \sum\limits_{k = 0}^{N_{\rm PE}} n^k_i\r)^2}{ \sum\limits_{j = 0}^{N_{\rm PE}} (n^k_i)^2} > 20\,,
\end{equation}
where the $n^k_i$ values are defined in \Cref{eq:Ii-icaro}.
For \chimera, instead, we checked that there were enough independent weighted samples to build the KDE by requiring 
\begin{equation}
     N^{\rm eff}_{\rm weights, i} = \f{\l( \sum\limits_{k = 0}^{N_{\rm PE}} w^k_i\r)^2}{ \sum\limits_{k = 0}^{N_{\rm PE}} (w^k_i)^2} > 2\;,
\end{equation}
where the $w^k_i$ values are given in \Cref{eq:weights-kde-chimera} and are normalized so that they sum to 1 for each event.

Finally, to estimate the detector selection function, we performed a large-scale injection campaign, drawing $5\times 10^7$ sources that span the whole detectable space, computed their S/N, and used the results to estimate the detection probability at each desired S/N cut.
The final number of injections exceeding the S/N threshold is $\sim 4.6\times10^6$ and $\sim 2.7\times10^6$ for S/N $>$ 60 and S/N $>$ 75, respectively, which is sufficient to ensure the numerical stability of the Monte Carlo integral in \Cref{eq:n-exp-approx}. For all three codes this is  \citep{Farr:2019rap, Talbot:2023pex, Heinzel:2025ogf}
\begin{equation}
    N^{\rm eff}_{\rm inj} = \f{\l[ \sum\limits_{j = 0}^{N_{\rm det}} s_j\r]^2}{\l[ \sum\limits_{j = 0}^{N_{\rm det}}s_j^2-N_{\mathrm{inj}}^{-1}\l( \sum\limits_{j = 0}^{N_{\rm det}} s_j\r)^2\r]} > 5 N_{\rm obs}\,, 
\end{equation}
where  the $s_j$ values are the terms in the sum over detected injections in \Cref{eq:n-exp-approx}.

\section{Results}
\label{sec:Results}

In this section we present the results obtained with the three pipelines and the two blinded mock BBH catalogs described in the previous section. 
This section is organized as follows.
In \Cref{subsec:code-performances} we analyze and compare the computational performance of the codes.
In \Cref{subsec:constraints} we present the constraints obtained from the S/N  >75 catalog using all three pipelines, and we show the improvement achieved by lowering the S/N threshold to 60, therefore including more events.
Finally, in \Cref{subsec:constraining-power} we discuss the characteristics---such as masses and redshifts---of the events producing the strongest constraints on the cosmological parameters.

To sample the likelihood implemented in \chimera we used the sequential Monte Carlo sampler \texttt{pocoMC} \citep{Karamanis:2022alw, Karamanis:2022ksp}. 
For the \pymcpop analyses, we employed the No-U-Turn Sampler (NUTS) within \texttt{PyMC}, which supports GPU sampling through the \texttt{numpyro} backend.
The priors adopted for each hyperparameter are shown in \Cref{tab:hyperparams_summary}.

\begin{table}[!ht]
\caption{Summary of hyperparameters and priors adopted. }
\centering\resizebox{\hsize}{!}{
\begin{tabular}{llc}
\hline
\hline
Symbol & Description &  Uniform prior boundaries \\
\hline
\multicolumn{3}{l}{Cosmology (flat $\Lambda$CDM)} \\
$H_0$ & Hubble constant $[\si{\kilo\meter\per\second\per\cubic\mega\parsec}]$ & $[20.0, 200.0]$ \\
$\Omega_{\rm m,0}$ & Matter energy density & $[0.01, 0.99]$ \\
\hline
\multicolumn{3}{l}{Mass distribution (Power Law + Gaussian Peak)} \\
$\alpha$ & Primary power law slope & $[1.5, 12]$ \\
$\beta$ & Secondary power law slope & $[-4, 12]$ \\
$\delta_m$ & Smoothing parameter $[\msun]$ & $[0.01, 10.0]$ \\
$m_{\rm low}$ & Power laws lower limit $[\msun]$ & $[2, 10]$ \\
$m_{\rm high}$ & Power laws upper limit $[\msun]$ & $[50, 200]$ \\
$\mu_{\rm g}$ & Gaussian peak position $[\msun]$ & $[20, 50]$ \\
$\sigma_{\rm g}$ & Gaussian peak width $[\msun]$ & $[0.4, 10]$ \\
$\lambda_{\rm g}$ & Gaussian peak weight & $[0.01, 0.99]$ \\
\hline
\multicolumn{3}{l}{Rate evolution (Madau-like)} \\
$\gamma$ & Slope at $z<z_p$ & $[0, 15]$ \\
$\kappa$ & Slope at $z>z_p$ & $[0, 20]$ \\
$z_{\rm p}$ & Peak redshift & $[0, 5]$ \\
\hline
$R_0$ & Local merger rate density $[\si{\per\cubic\mega\parsec\per\years}]$ &  $[0.01,100]$ \\
\hline
\end{tabular}
}
\label{tab:hyperparams_summary}
\end{table}

\subsection{Code performance}\label{subsec:code-performances}
To profile the likelihoods of \chimera and \icaro, we tracked the time required for a single likelihood evaluation on a NVIDIA Ampere A100 GPU with 64\,GB RAM. 
This setup enables a direct performance comparison between the two codes.  
The computational times are averaged over 1000 different sets of cosmological and population parameters drawn from their posterior distribution described in the following section.
For the S/N $>$ 75 catalog, we find that \icaro requires $0.093\,\mathrm{s}$ per likelihood evaluation, while \chimera requires $0.037\,\mathrm{s}$, making the latter $2.5$ times faster. 
For the S/N $>$ 60 catalog, the average evaluation time becomes $0.064\,\mathrm{s}$ for \chimera and $0.14\,\mathrm{s}$ for \icaro, reducing the speed-up factor of \chimera to $2.2$. 
Because \pymcpop is sampled with the NUTS sampler, the relevant unit of computational work is a Hamiltonian integration iteration rather than a single evaluation of the likelihood.
We therefore track both the wall-time per iteration and the typical number of leapfrog steps performed within each NUTS trajectory.
We compiled the model on GPU using the JAX/\texttt{numpyro} backend and ran two short chains of 30 iterations to probe the performance.
For the S/N $>$ 75 catalog, the average cost is $\simeq 22.6\,\mathrm{s/it}$, with $N_{\rm step}=1023$ leapfrog steps per iteration.
Therefore, the effective cost per leapfrog step is $t_{\rm step}\simeq 22.6/1023 \simeq 0.022\,\mathrm{s}$.
Since each leapfrog step requires one new evaluation of the likelihood and its gradient, this estimate provides a direct proxy for the effective $(\log\mathcal{L}+\nabla\log\mathcal{L})$ cost during sampling. 
For the S/N $>$ 60 catalog, we obtained an average wall-time of $\simeq 30.2\,\mathrm{s/it}$, with a typical NUTS trajectory length of $N_{\rm step}=1023$ leapfrog steps.
The effective cost in this case is therefore $t_{\rm step}\simeq 30.2/1023 \simeq 0.03\,\mathrm{s}$ per leapfrog step, i.e., per compiled $(\log\mathcal{L}+\nabla\log\mathcal{L})$ evaluation.

The population fit for the S/N $>$ 75 catalog using \texttt{pocoMC} required approximately $6\times10^5$ likelihood calls and $15.5$ GPU hours on a single A100 GPU for \icaro, while the fit with \chimera used $\sim6.8\times10^5$ likelihood calls and about $7$ GPU hours. 
For the S/N $>$ 60 catalog, the required resources were $19$ GPU hours ($4.6\times10^5$ likelihood calls) for \icaro and $11.9$ GPU hours ($6.7\times10^5$ likelihood calls) for \chimera.
The total GPU time for a NUTS MCMC run with \pymcpop is computed as $t_{\rm step} \times N_{\rm step} \times (N_{\rm tune}+N_{\rm draws})\times N_{\rm chains}$. 
Here $N_{\rm tune}$ is the number of steps in the warm-up phase (typically 500 for  simulations or 1000 for real data), and $N_{\rm draws}$ is the number of sampled points per chain (typically 1000 with simulations). 
The $N_{\rm chains}$ factor depends on the device. If GPU memory allows, one can vectorize the computation to reduce this factor; otherwise, one has to do sequential sampling chain by chain.
Assuming $N_{\rm chains} = 4$, the MCMC fit of the S/N $>$ 75 and S/N $>$ 60 catalogs with \pymcpop would require approximately 37.5 and 51 GPU hours, respectively.

We summarize the single-evaluation times and the convergence times in the bottom and top panels of \Cref{fig:computational-times}, respectively.
The timing results for the two analyzed catalogs are shown as points connected by a solid line.

\begin{figure}[!htb]
    \centering
    \resizebox{\hsize}{!}{\includegraphics[width=\textwidth]{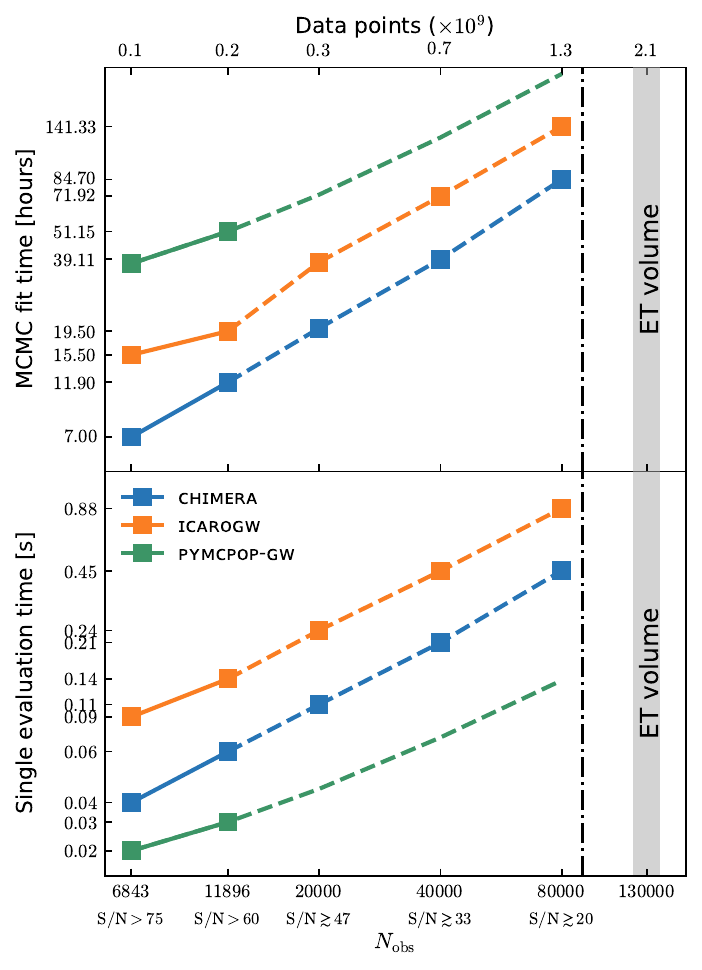}}
    \caption{Scaling of computational times for a single evaluation and for a complete population fit as a function of the number of events and the total number of data points $D = 3\times(N_{\rm obs}\times N_{\rm PE} + N_{\rm inj})$, where $N_{\rm PE}=5000$ per event and $N_{\rm inj} = N_{\rm obs}\times 390$. Markers connected by a solid line correspond to actual simulated catalogs with S/N cuts of 60 and 75, as labeled. Markers connected by a dashed line refer to timings obtained using projected catalogs constructed by stacking events and injections from the S/N $>$ 60 catalog.
    The corresponding S/N cut for this number of events is indicated in the $x$-axis. Black vertical lines show the memory saturation limits for \chimera. The gray bands indicate the approximate event volume expected for ET in one year of observations.}
    \label{fig:computational-times}
\end{figure}

In \Cref{fig:computational-times}, we also forecast how the computational times for a single evaluation and for a population fit scale with the number of events and the total dataset size.
These results are obtained by stacking a set of events from the S/N $>$ 60 catalog until they reach a total number of events, $N_{\rm ev}$, equal to $2\times10^4$, $4\times10^4$, and $8\times10^4$.
These correspond to approximate S/N cuts in our population of $47$, $33$, and $20$, respectively.
The total number of data points, $D$, for each case is $D=3\times(N_{\rm ev}\times N_{\rm PE} + N_{\rm inj})$, where the factor 3 represents the number of GW event parameters, i.e., the two binary masses and the luminosity distance.
In each case, we consider $N_{\rm PE} = 5000$ per event, and we scale the number of injections as $N_{\rm inj} = N_{\rm ev}\times 390$.
The factor $390$ was chosen because it matches the value present in the S/N $>$ 60 catalog and is usually sufficient to ensure enough effective injections for the numerical stability of the Monte Carlo sum in \Cref{eq:n-exp-approx}.
The resulting times for \chimera and \icaro are shown in \Cref{fig:computational-times} with markers connected by a dashed line.
We observe a perfectly linear scaling with the number of events (or the total number of data points) for both \chimera and \icaro. 
The total MCMC fit times shown in the top panel of \Cref{fig:computational-times} were instead obtained by multiplying the single-evaluation time by the number of likelihood calls required for convergence, which can vary between runs. 
In particular, for \icaro, the projected MCMC times are conservatively computed assuming the maximum number of likelihood calls observed across the two measured catalogs.
We also note a constant speedup factor of $1.5$--$2$ for \chimera compared to \icaro.
On the other hand, \chimera saturates the RAM earlier than \icaro.
Specifically, under our test conditions, \chimera cannot process more than $1\times10^5$ events (black vertical line in \Cref{fig:computational-times}), while \icaro can handle at least the same number of events without memory issues.
Overall, a single GPU can fit a large fraction of the total volume of data predicted for ET (gray band in \Cref{fig:computational-times}), which is about $10^5$ events.

We do not forecast timings for \pymcpop, but instead plot the linear fit of the two measured points, justified by the linear scaling observed for the other codes and noted for \pymcpop at lower event counts.
The reason is that, in its current development stage, \pymcpop is unable to produce fully converged chains on the GPU for the full data volumes considered in this work.
Although the method has been validated on smaller datasets ($\sim10^3$ events), scaling it to the number of events analyzed here caused GPU memory and accuracy problems.
In particular, directly materializing the full injection set for the Monte Carlo integral in \Cref{eq:n-exp-approx} triggers out-of-memory errors on the GPU used in our tests. 
To overcome this, \pymcpop implements a batched estimator for $N_{\rm exp}$, where injections are loaded and processed in chunks of $2\times10^4$ injections.
Despite this memory optimization, GPU sampling on the full mock catalogs still exhibits poor chain mixing after thousands of iterations. As a result, the converged \pymcpop posterior samples presented in this paper were obtained via CPU runs. 
We attribute this remaining limitation to numerical aspects affecting gradient-based sampling at scale (e.g., interpolant resolution and the smoothing of sharp prior boundaries), which are being actively improved. 
Nonetheless, our tests confirm that the batched Monte Carlo strategy is fundamentally capable of supporting the data volumes expected in the ET era.

We note that all reported computation times for \chimera and \icaro were obtained using 5000 PE samples per event in each catalog.
While this number is sufficient to approximate the mock GW posterior distributions obtained in this work using a FIM approach, real events may require more PE samples to accurately represent the single-event posterior, which would modify the computational times reported previously. In particular, if stricter numerical stability estimators for the likelihood \cite{Heinzel:2025ogf} are considered.
To test this, we performed a population fit of the S/N $>$ 60 catalog with 10000 PE samples per event using \chimera. 
With this larger catalog a single likelihood evaluation takes $0.11\,\mathrm{s}$, and the complete population fit requires $22$ GPU hours.
This time is similar to the case with $2\times10^4$ events, indicating that the time scaling shown in \Cref{fig:computational-times} is a function of the number of data points passed to the inference.

We have demonstrated that, on a single GPU, these GPU-ready and optimized pipelines can process up to  $10^5$ events---with up to $10^4$ PE samples per event---and \(2.7\times10^6\) injections in about one or two weeks of dedicated computing time. 
Since the likelihood in \Cref{eq:hierarchical-lkl} is perfectly parallelizable over events, meaning the single event contribution to the likelihood can be computed independently and combined at the end, the optimal strategy for the ET era is to exploit this property by distributing the workload across multiple GPUs. 
This can be achieved using well-established protocols such as the message passing interface (MPI) or through new JAX sharding features \citep{jax2018github}. 
By splitting the dataset across a few GPUs, a number already accessible at current HPC facilities, pipelines can easily scale to handle the volume of data of the ET. 
However, we caution that more complex astrophysical models or the requirement of stricter numerical stability estimators might require more memory.
Such a demand may be prohibitive for the expected volume of GW detections from ET. 
Therefore, further computational improvements to \icaro, \pymcpop, and \chimera may be necessary. 

\subsection{Constraints}\label{subsec:constraints}

\begin{figure*}[!htb]
    \centering
    \includegraphics[width=0.95\textwidth]{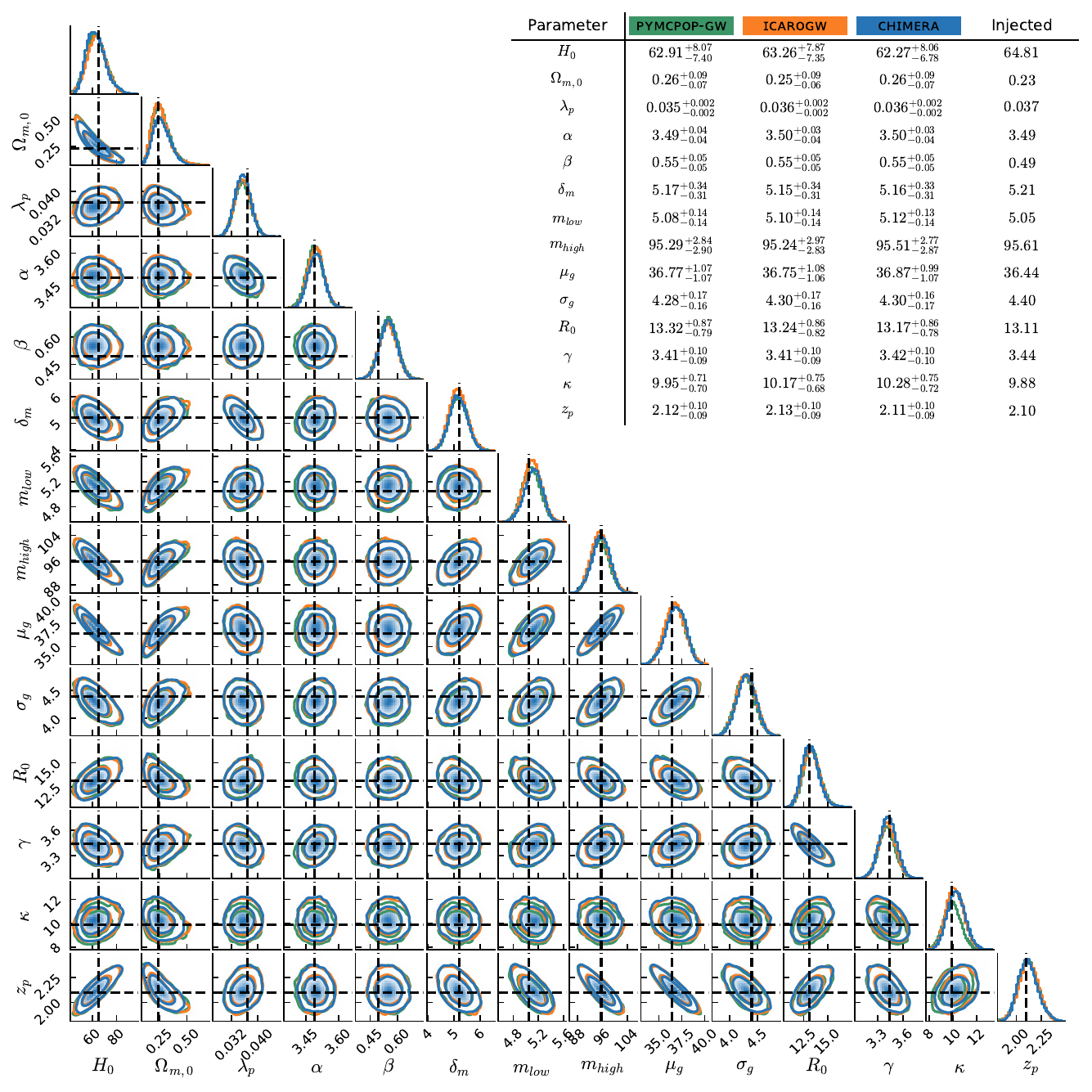}
    \caption{Cosmological and population parameter constraints from the S/N $>$ 75 catalog. The corner plot shows 1D marginalized posteriors and 2D contours (68\% and 95\% credible regions) obtained with all three pipelines. The inset  provides a table of the median values and the 68\% C.I. for each parameter.
    }
    \label{fig:corner-snr75}
\end{figure*}

\noindent Figure \ref{fig:corner-snr75} presents a corner plot showing the 1D marginalized posterior distributions and 2D marginalized contours at the 68\% and 95\% credible regions for all the population and cosmological parameters obtained with the S/N $>$ 75 catalog.
The table in the inset summarizes the median values and 68\% credible intervals (C.I.) for each parameter as obtained via all three codes.

The constraints from all three pipelines are in excellent agreement; each recovers fully compatible contours, and exhibits the same correlation patterns among cosmological and population parameters.
All three codes recover nearly every fiducial cosmological and population parameter within the 68\% C.I., and every parameter within the 95\% C.I.
Thanks to the depth and size of the ET catalog, we can reveal for the first time correlations between the cosmological parameters $H_0$, $\Omega_{\mathrm{m},0}$ and the population parameters, which have not been observed in studies currently available.
For example, we find that the cosmological parameters correlate with all the population parameters related to mass features, while in current constraints the only correlation that we can observe is the one with the location of two mass features at $\sim 10 M_{\odot}$ and $\sim 30 M_{\odot}$ in the BBH mass spectrum \citep{LIGOScientific:2025jau}.
We also note that because the catalog extends to redshifts beyond the peak of the GW merger rate $z_p$, modeled here following the Madau--Dickinson star formation rate, we can directly measure $z_p$ rather than placing only lower bounds as with the current redshift-limited data.
This is also of crucial importance for dark siren cosmology, as the position of $z_p$ provides another population-level redshift scale that can be used to infer cosmology \citep{Ye:2021klk}.
This demonstrates the importance of the ET for constraining and understanding the population of BBHs, particularly at high redshift. 

A central physical problem beyond the scope of this work, which instead focuses more on the analysis feasibility, is the impact that a redshift-evolving BBH distribution would have on the cosmological inference. 
Recent studies \citep{Mukherjee:2021rtw, Karathanasis:2022rtr, Pierra:2023deu, Agarwal:2024hld} argue that a redshift evolution of the mass spectrum could already bias the cosmological parameters inference for current-generation GW events. 
Possible mechanisms that could source redshift evolution in the mass spectrum---or contaminate it in general---include contributions from multiple formation channels \citep{Wong:2020ise,Zevin:2022bfa, Li:2023yyt,  Torniamenti:2024uxl,Fishbach:2024hfw, Li:2024rmi}, Population III star mergers \citep{Woosley:2021xba,vanSon:2021zpk}, mergers in dense stellar clusters \citep{Antonini:2016gqe,Mapelli:2021gyv,Gerosa:2021mno,Ye:2024ypm}, or mergers in AGN accretion disks \citep{Ford:2021kcw}.
This problem will be even more important for ET, given its extended redshift range.
A viable approach to incorporating redshift evolution in the mass spectrum without biasing cosmological analyses may rely on the use on nonparametric models, such as the ones proposed in \citet{Rinaldi:2023bbd, Heinzel:2024hva, Farah:2024xub, Sadiq:2025aog}.

\begin{figure*}
    \centering
    \includegraphics[width=0.95\textwidth]{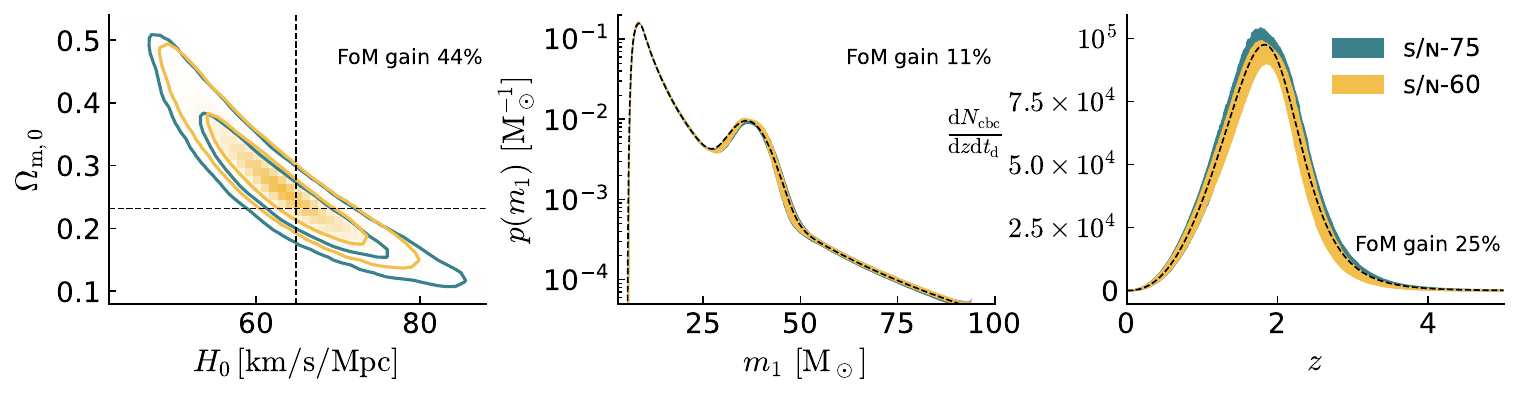}
    \caption{Comparison of constraints from the S/N $>$ 60 and S/N $>$ 75 catalogs. Left: 2D marginalized posterior for $(H_0, \Omega_{\rm m,0})$. In particular, we plot the 68\% and 95\% credible regions. Center and right: Predictive posterior distributions for the primary mass spectrum and the redshift event rate \Cref{eq:p-z}, respectively. The dashed lines indicate the blinded fiducial values and population model.}
    \label{fig:results-snr60-vs-snr75}
\end{figure*}

To assess how the inclusion of lower S/N events affects the inference, we compare the constraints obtained from the S/N $>$ 60 and S/N $>$ 75 catalogs.
This comparison is performed by combining the samples obtained from the three different pipelines for each catalog.
Figure \ref{fig:results-snr60-vs-snr75} presents this comparison across three different metrics: the joint constraints on the Hubble constant and matter density parameter $(H_0, \Omega_{\mathrm{m},0})$; the predictive posterior distribution of the primary mass spectrum $p(m_1)$; and the event rate distribution per unit redshift and detector-frame time \Cref{eq:p-z}.
To quantify the improvement in these three metrics, we also computed their figure of merit (FoM), defined as the inverse of the area of the 68\% credible region.
The improvement factors (see insets in  \Cref{fig:results-snr60-vs-snr75}) are significant for all three metrics.
Whereas the improvements in the $H_0$-$\Omega_{\mathrm{m},0}$ constraints and in the rate distribution $\frac{\dd N_{\mathrm{cbc}}}{\dd t_\mathrm{d} \dd z}$ are visually apparent in the figure, the FoM is essential for assessing the improvement in the mass distribution. 
In fact, the two reconstructed curves for $p(m_1)$ are very similar, as both are precise and close to the fiducial distribution.

We find a joint precision of 12\% on $H_0$ and 31\% on $\Omega_{\rm m,0}$ for the S/N $>$ 75 catalog,  which contains 6843 BBHs.
Although the S/N $>$ 60 catalog includes approximately twice the number of events, the constraints on $H_0$ and $\Omega_{\mathrm{m},0}$ remain at the 10.5\% and 26\% level, respectively. 
The improvement is consistent with a $\sim\sqrt{2}$ improvement given by the increased number of GW events considered in the analysis.
In addition to the parameters of the distance--redshift relation, spectral sirens can yield a valuable constraint on the Hubble function $H(z)$ at intermediate redshifts, provided a $\Lambda$CDM evolution is assumed \citep{Farr:2019twy,Mancarella:2021ecn,Farah:2024xub}.  
To prove this, we compute $H(z)/(1+z)$ on a redshift grid for each sample of $H_0$ and $\Omega_{m,0}$, thereby obtaining samples that represent the posterior predictive distribution of $H(z)/(1+z)$. From this distribution, we then calculate the median and the 68\% C.I.
Figure \ref{fig:Hz-ppd} shows this constraint (top), together with the relative precision (bottom).
For the S/N $>$ 75 and S/N $>$ 60 catalogs, we find that the best constraints on $H(z)$ are achieved at redshifts $z \approx 1.41$ and $z \approx 1.53$ (vertical dashed lines in the bottom panel of \Cref{fig:Hz-ppd}), with a corresponding precisions of 3.04\% and 2.37\%, respectively.
Interestingly, the redshift range where we get the best constraints on $H(z)$ is close to the one typically used to constrain the transition redshift of the Universe’s expansion history \citep{Moresco:2016mzx}.
For example, in the $0.7 < z < 1.8$ interval, the mean precisions on $H(z)$ are 3.5\% (S/N $>$ 75) and 2.8\% (S/N $>$ 60).
Having the tightest spectral siren constraints in this interval is promising, since estimates of $H(z)$ in this range usually rely on not fully cosmology-independent cosmological probes.
We note, however, that the strength of the constraint might depend on assumptions on the fiducial cosmological model \citep{Pierra:2025hoc}.

These findings differ from those of \citet{Califano:2025qbx}, who reported a $\sim6\%$ measurement on both parameters using only 1659 events above S/N > 80 (though assuming the ET in a triangular configuration).
This difference is also notable because the predictive posterior distributions of the mass and redshift rate distributions shown in \Cref{fig:results-snr60-vs-snr75} are considerably tighter than those presented by this analysis, as would be expected given our significantly larger event sample.
Additionally, our analysis reveals a distinct set of correlations between cosmological and population parameters that were not identified in previous studies \citep{Califano:2025qbx}.
Further, we also mention that \cite{Chen:2024gdn} found that with 3G detectors it will be possible to achieve percent-level constraints on the Hubble constant and $\Omega_{\rm m, 0}$ using $5\times 10^4$ events.
Rescaling our result to the same number of events using the expected statistical scaling $1/\sqrt{N}$ yields an uncertainty of order $\sim 4\%$. 
From Fig.~6 of \cite{Chen:2024gdn}, one can infer an uncertainty of $\sim$$2\%$ on $H_0$, which is broadly comparable once the methodological and modeling differences between the two analyses are taken into account.\footnote{In addition,  the $1/\sqrt{N}$ scaling does not account for the different S/N cuts in the two analyses.}
Two aspects are particularly worth noting. 
First, single-event error estimates in \citet{Chen:2024gdn} are calibrated on simulations of CE performed several years ago \citep{Vitale:2016icu}, and sensitivity curves are known to evolve over time as detector designs are updated. 
Second, their population model includes binary neutron star systems, which provide an additional and important mass scale that strengthens the constraints \citep{Ezquiaga:2022zkx}. 
Overall, these considerations suggest that achieving percent-level constraints on $H_0$ with spectral sirens is possible with (1) as large a number of events as possible, rather than restricting the analysis to only high S/N detections, and (2) the inclusion of as many mass scales as possible in the observed population. 
A further conclusion, consistent with previous work, is that the strongest constraints from spectral sirens are not obtained on $H_0$, but rather on the expansion rate $H(z)$ at intermediate redshifts. 
Interestingly, our findings suggest that this is possible even with a smaller sample of high S/N events (see \Cref{fig:Hz-ppd}).

\begin{figure}
    \centering
    \includegraphics[width=0.95\linewidth]{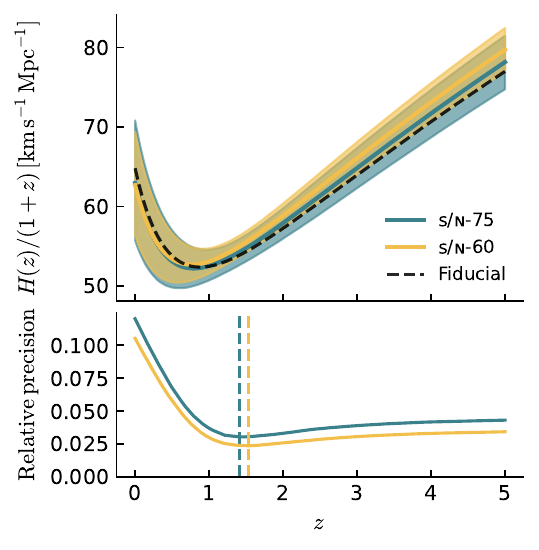}
    \caption{Top:  1$\sigma$ contours of the predictive posterior distribution for $H(z)/(1+z)$. Bottom: Relative precision on $H(z)$, computed as the width of the 1$\sigma$ C.I. divided by twice the median of $H(z)$. The dashed vertical lines indicate the redshift at which the constraint on $H(z)$ is strongest.}
    \label{fig:Hz-ppd}
\end{figure}

While the tighter constraints on $H_0$, $\Omega_{\mathrm{m},0}$ reported by earlier works may be due to the use of population models with sharper features than the blinded one adopted in this study \citep[e.g., in][]{Califano:2025qbx,Chen:2024gdn} or to the larger dataset considered \citep{Chen:2024gdn}, this comparison underscores the need for fiducial forecasts in the era of the ET and third-generation detectors.
Such forecasts should be based on realistic catalogs generated beyond the Fisher-matrix approximation and should incorporate updated population models and sensitivity curves to properly assess the cosmological potential of ET using spectral sirens.
As this detailed forecast study was beyond the scope of the present work, we defer it to future investigations.
If the results found in this work are revealed to be accurate, achieving constraints on $H_0$ comparable to those from the cosmic microwave background and supernovae with just high S/N events from 3G detectors may require the complementary use of galaxy catalogs, sharper features in the population model~\citep{LIGOScientific:2025pvj, Tiwari:2025oah, Pierra:2026ffj, Tagliazucchi:2026gxn}, or additional information on $\Omega_{\rm m,0}$ that can be obtained from other probes such as baryonic acoustic oscillations \citep{DESI:2024mwx} or uncalibrated supernova measures \citep{Riess:2021jrx}. 
In fact, fixing $\Omega_{\rm m,0}$ to its fiducial value improves the constraint on $H_0$ to 3\% using the S/N $>$ 75 catalog, corresponding to an improvement of a factor of $4$  over the case where $\Omega_{\rm m,0}$ is marginalized over (see \Cref{app:Om0fixed-case}).
Otherwise, the inclusion of a larger sample of spectral sirens can be envisaged, with the corresponding accurate treatment of selection effects. A valuable constraint on $H(z)$ at intermediate redshifts seems a concrete possibility in all cases.

\subsection{Constraining power}\label{subsec:constraining-power}

We assessed which events contribute most significantly to constraining cosmological parameters and population hyperparameters, focusing specifically on $H_0$, $\Omega_{\rm m,0}$, and $\mu_g$. 
To quantify each event's contribution, we computed Pearson correlation coefficients, $\rho_{i,\lambda}$, between the marginal likelihood of the $i$-th event and the aforementioned hyperparameters.
The $i$-th event marginal likelihood is the single-event contribution to the total hyperlikelihood \Cref{eq:hierarchical-lkl}, i.e.,
\begin{equation}
    \log\mathcal{L}_i(\dd_i \mid \boldsymbol{\Lambda}) \propto  \log\l(\!\int\! \dd \boldsymbol{\theta}_{\dd,i}  \mathcal{L}_{\rm gw}\l( \boldsymbol{d}_i  \condmid  \boldsymbol{\theta}_{\dd,i}\r) \left| \f{\dd \boldsymbol{\theta}_i}{\dd \boldsymbol{\theta}_{\dd,i}} \right| \f{\dd N}{\dd \boldsymbol{\theta}_i}(\boldsymbol{\Lambda})\r) - \f{N_{\rm exp}(\boldsymbol{\Lambda})}{N_{\rm obs}}.
\end{equation}
The resulting coefficients are shown in \Cref{fig:constraining-power}, where we show the 2D histogram of the mock GW events with S/N $>$ 75 in the $m_{\dd,1}$-$d_L$ plane.
The color of each $(30\times30)$ logarithmically spaced bin represents the mean Pearson correlation coefficient calculated from the events falling within that bin.
The curves overlaid on each panel correspond to $m_{\dd,1}\cdot p(m_1(m_{\dd,1}, d_L, \boldsymbol{\lambda}_c) \mid \boldsymbol{\lambda}_m)$ at different values of $H_0$ (but with $\boldsymbol{\lambda}_m$ fixed to the fiducial) and at various luminosity-distance levels. 
In particular, we plot these curves for $d_L = 1,\,2,\,6,\,\mathrm{and}\,16\,\si{\giga\parsec}$, corresponding to $z = 0.19,\,0.34,\,0.84,\,\mathrm{and}\,1.83$ given the fiducial $H_0\approx 64.8\;\si{\kilo\meter\per\second\per\cubic\mega\parsec}$.

\begin{figure*}
    \centering
    \includegraphics[width=0.95\textwidth]{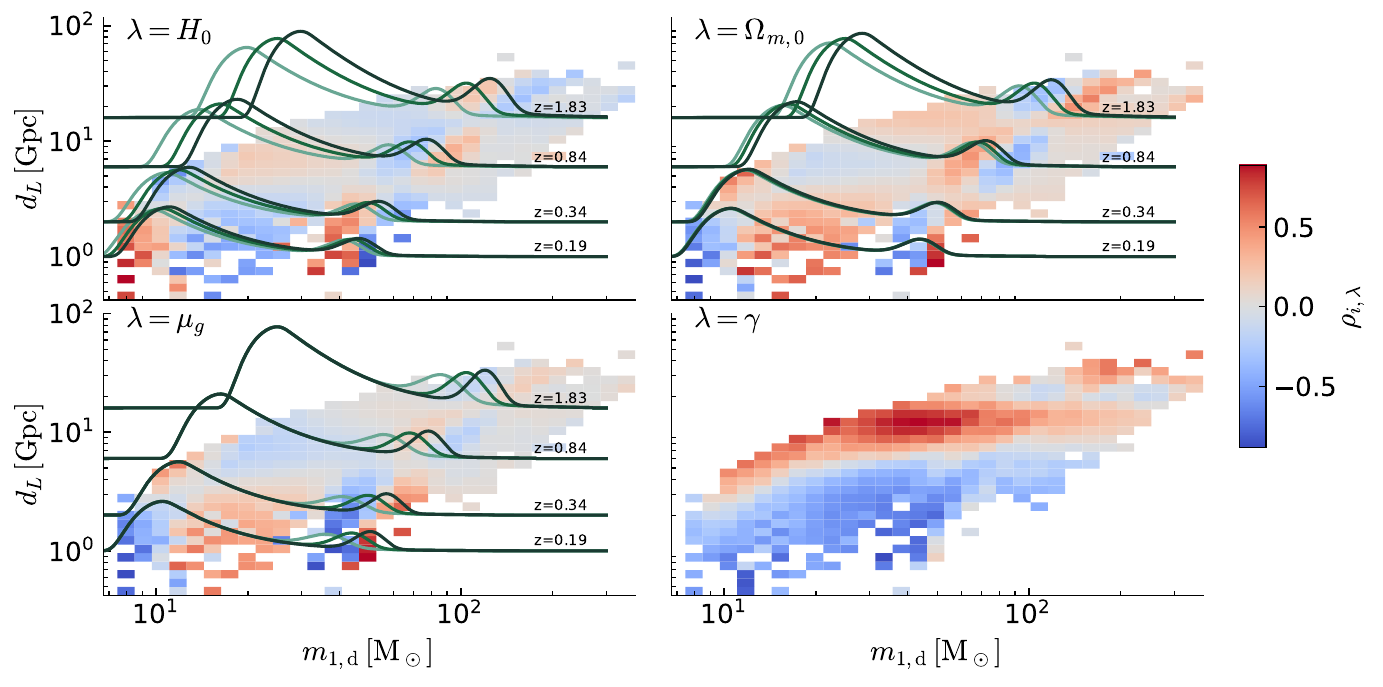}
    \caption{2D histogram of the events with S/N $>$ 75 in the $d_L$-$m_{\dd,1}$ plane. The color-coding indicates the mean Pearson correlation coefficient between the marginal likelihood of events in each bin and the hyperparameters $H_0$ (top left), $\Omega_{\mathrm{m},0}$ (top right), $\mu_g$ (bottom left), and $\gamma$ (bottom right). 
    In each panel we plot the function $m_{\dd,1}\cdot p(m_{\dd,1})$ for different values of the hyperparameter considered and at various luminosity-distance levels. In particular, the curves are calculated for (from light to dark green): $H_0 \in [40, 64.8, 90]\si{\kilo\meter\per\second\per\mega\parsec}$, $\Omega_{\rm m,0} \in [0.05, 0.23, 0.55]$, and $\mu_g \in [30, 36.44, 42] \msun$.     The redshifts corresponding to the distances considered (calculated with the fiducial $H_0\approx 64.8\;\si{\kilo\meter\per\second\per\mega\parsec}$) are also indicated.}
    \label{fig:constraining-power}
\end{figure*}

The events with the highest Pearson coefficients are those that constrain the considered hyperparameters most strongly.
For $H_0$, $\Omega_{\rm m,0}$, and $\mu_g$ these are found at low luminosity distances ($d_L < 2\,\si{\giga\parsec}$) and around the structures of the mass spectrum, while for $\gamma$ they are distributed across the whole parameter space.
In particular, events with masses lower than the Gaussian peak are negatively correlated with $\mu_g$, whereas those with masses higher than the peak are positively correlated.
This means that events below and above the peak pull $\mu_g$ toward lower and higher values, respectively, with the final constraint set by the balance of these two effects.
Similarly, constraints on  $H_0$ and $\Omega_{\mathrm{m},0}$ are driven mainly by events near the low-mass edge ($m_{\rm low}$) and around the Gaussian peak.
For $H_0$, the correlation signs near the peak are opposite to those for $\mu_g$, consistent with the $H_0$-$\mu_g$ anticorrelation visible in \Cref{fig:corner-snr75}.
The physical origin of this anticorrelation is that a higher $H_0$ increases the redshift corresponding to a fixed $d_L$, which in turn requires a smaller inferred $\mu_{\rm g}$ to match the observed data. 
The correlation-sign pattern for $\Omega_{\mathrm{m},0}$ is opposite to that for $H_0$, reflecting their mutual anticorrelation (see \Cref{fig:results-snr60-vs-snr75}).

Regarding $\gamma$, low-distance events tend to reduce its value, while events with $d_L > 6\,\si{\giga\parsec}$ (corresponding to $z\approx 0.84$ in the fiducial cosmology, near the peak in the observed redshift distribution; \Cref{fig:cat-props}) favor a higher value of $\gamma$.  
Moreover, around this luminosity distance value the correlation-sign pattern for $H_0$, $\Omega_{\rm m,0}$, and $\mu_g$ reverses.
This occurs because at these redshifts events constrain mainly $\Omega_{\rm m,0}$ rather than $H_0$: the Pearson coefficients relative to $H_0$ and $\mu_g$ become smaller, whereas those relative to $\Omega_{\mathrm{m},0}$ remain relevant.
Since $H_0$ and $\Omega_{\rm m,0}$ are anticorrelated, the sign pattern flips for the three hyperparameters.
To verify this, we computed the Pearson coefficients between events and $H_0$, $\mu_g$, and  $\gamma$, while fixing $\Omega_{\rm m,0}$ to its fiducial value.
The resulting coefficients, shown in \Cref{app:Om0fixed-case}, confirm that the sign reversal does not occur in that case at this distance.

At even greater distances---around $d_L\approx 16\,\si{\giga\parsec}$ ($z\approx 1.83$)---events primarily constrain $\Omega_{\mathrm{m},0}$ and $\gamma$.
Furthermore, the correlation-sign pattern for $H_0$, $\Omega_{\rm m,0}$, and $\mu_g$ reverts once more to that seen at lower luminosity distances.
This second transition occurs because, at this distance, the correlation of $\gamma$ with events that lie near the peak of the true redshift distribution changes sign.
Since $\gamma$ itself is correlated with $\Omega_{\rm m,0}$, $\mu_g$ and anticorrelated with $H_0$, the change in how events influence $\gamma$ propagates and reverses the sign patterns for the other hyperparameters.
This interpretation is supported by the case in which $\Omega_{\rm m,0}$ is fixed (see \Cref{app:Om0fixed-case}), where the same transition is seen at the same luminosity distance.

\section{Conclusion} 
\label{sec:Conclusion}

In this work we performed a blinded mock data challenge to test current spectral siren cosmological inference pipelines for the high-statistics regime of 3G interferometers.
We focused on assessing the computational scalability, the agreement between independent code implementations, and the constraining power for cosmology expected from an observatory such as the ET.
Our study yields several key conclusions. 

First, we demonstrated that GPU-accelerated pipelines, such as \icaro, \chimera, and \pymcpop, can handle the large data volume expected from third-generation detectors. 
By profiling their performance, we found that these codes can process up to $\sim 8\times10^{4}$--$10^5$ events with thousands of PE samples each on a single GPU within a manageable timeframe, confirming that GPU acceleration is the key for cosmological analyses in the third-generation era.
Second, through a blinded comparison, we validated that three independent implementations of the spectral siren likelihood produce fully consistent cosmological and population parameter constraints, confirming the robustness of the method.
Third, we forecast joint cosmological and population parameter constraints from the blinded population model, finding a 2.37\% constraint on the Universe's expansion history $H(z)$ at $z\approx1.53$ using $11896$ events detected at high S/N by the ET.
This corresponds to joint constraints of $10.5\%$ on $H_0$ and $26.5\%$ on $\Omega_{\mathrm{m},0}$.
Moreover, the mean precisions on $H(z)$ in the $0.7<z<1.8$ range are 3.5\% for the S/N $>$ 75 catalog and 2.8\% for the S/N $>$ 60 catalog.
Finally, we identified the specific events that drive cosmological constraints, revealing distinct correlation patterns between hyperparameters and features in the $m_{\dd,1}-d_L$ distribution of events.
In particular, we found that low-distance events constrain well all the cosmological parameters, while at larger distances the events constrain mainly $\Omega_{\mathrm{m},0}$.

Looking ahead, several important directions emerge from this study.
While GPU acceleration already enables the analysis of a large fraction of events that will be detected by the ET, analyzing future catalogs entirely will require the pipelines to evolve further. 
Key developments include advanced parallelization strategies using distributed GPU computing and efficient GPU memory handling.
Our analysis assumed a simple non-evolving population model with randomly chosen fiducial hyperparameters. Future work must incorporate more realistic astrophysics to provide more robust cosmological constraints.
Since ET will detect events at higher redshifts, and since spectral sirens can be combined with galaxy catalogs or additional probes (e.g., BAO) to increase their constraining power, future cosmological forecasts should also include  the evolving dark energy equation of state. 

Nevertheless, our results confirm the feasibility of spectral siren cosmology with a large number of events. 
They provide a validated, performance-tested framework that paves the way for precision cosmology with third-generation GW observatories.

\begin{acknowledgements}
    We acknowledge the ICSC for awarding this project access to the EuroHPC supercomputer LEONARDO, hosted by CINECA (Italy).
    M.T. acknowledges the funding from the European Union - NextGenerationEU, in the framework of the HPC project – “National Center for HPC, Big Data and Quantum Computing” (PNRR - M4C2 - I1.4 - CN00000013 – CUP J33C22001170001). M.Mo. acknowledges the financial contribution from the grant PRIN-MUR 2022 2022NY2ZRS 001 “Optimizing the extraction of cosmological information from Large Scale Structure analysis in view of the next large spectroscopic surveys”. M.Mo. and N.B. acknowledge support from the grant ASI n. 2024-10-HH.0 “Attività scientifiche per la missione Euclid – fase E”.
    The work of M.Ma. and A.A. is supported by the French government under the France 2030 investment plan, as part of the Initiative d'Excellence d'Aix-Marseille Universit\'e -- A*MIDEX AMX-22-CEI-02.
    S.M. is supported by ERC grant GravitySirens  101163912. Funded by the European Union. Views and opinions expressed are however those of the author(s) only and do not necessarily reflect those of the European Union or the European Research Council Executive Agency. Neither the European Union nor the granting authority can be held responsible for them.
    S.F. work received support from the French government under the France 2030 investment plan, as part of the Excellence Initiative of Aix Marseille University - amidex (AMX-19-IET-008 -IPhU). 
\end{acknowledgements}

\bibliographystyle{aa} 
\bibliography{ref.bib}

@article{LIGOScientific:2016aoc,
    author = "Abbott, B. P. and Abbott, R. and Abbott, T. D. and others",
    collaboration = "LIGO Scientific, Virgo",
    title = "{Observation of Gravitational Waves from a Binary Black Hole Merger}",
    eprint = "1602.03837",
    archivePrefix = "arXiv",
    primaryClass = "gr-qc",
    reportNumber = "LIGO-P150914",
    doi = "10.1103/PhysRevLett.116.061102",
    journal = "Phys. Rev. Lett.",
    volume = "116",
    number = "6",
    pages = "061102",
    year = "2016"
}

@article{Schutz:1986gp,
    author = "Schutz, Bernard F.",
    title = "{Determining the Hubble Constant from Gravitational Wave Observations}",
    doi = "10.1038/323310a0",
    journal = "Nature",
    volume = "323",
    pages = "310--311",
    year = "1986"
}

@article{Holz:2005df,
    author = "Holz, Daniel E. and Hughes, Scott A.",
    title = "{Using gravitational-wave standard sirens}",
    eprint = "astro-ph/0504616",
    archivePrefix = "arXiv",
    doi = "10.1086/431341",
    journal = "ApJ",
    volume = "629",
    pages = "15--22",
    year = "2005"
}

@article{Moresco:2022phi,
    author = "Moresco, Michele and Amati, Lorenzo and Amendola, Luca and others",
    title = "{Unveiling the Universe with emerging cosmological probes}",
    eprint = "2201.07241",
    archivePrefix = "arXiv",
    primaryClass = "astro-ph.CO",
    doi = "10.1007/s41114-022-00040-z",
    journal = "Living Rev. Rel.",
    volume = "25",
    number = "1",
    pages = "6",
    year = "2022"
}

@article{Jin:2025dvf,
    author = "Jin, Shang-Jie and Song, Ji-Yu and Sun, Tian-Yang and Xiao, Si-Ren and Wang, He and Wang, Ling-Feng and Zhang, Jing-Fei and Zhang, Xin",
    title = "{Gravitational wave standard sirens: A brief review of cosmological parameter estimation}",
    eprint = "2507.12965",
    archivePrefix = "arXiv",
    primaryClass = "astro-ph.CO",
    doi = "10.1007/s11433-025-2829-9",
    journal = "Sci. China Phys. Mech. Astron.",
    volume = "69",
    number = "2",
    pages = "220401",
    year = "2026"
}

@article{DelPozzo:2011vcw,
    author = "Del Pozzo, Walter",
    title = "{Inference of the cosmological parameters from gravitational waves: application to second generation interferometers}",
    eprint = "1108.1317",
    archivePrefix = "arXiv",
    primaryClass = "astro-ph.CO",
    doi = "10.1103/PhysRevD.86.043011",
    journal = "Phys. Rev. D",
    volume = "86",
    pages = "043011",
    year = "2012"
}

@article{Chen:2017rfc,
    author = "Chen, Hsin-Yu and Fishbach, Maya and Holz, Daniel E.",
    title = "{A two per cent Hubble constant measurement from standard sirens within five years}",
    eprint = "1712.06531",
    archivePrefix = "arXiv",
    primaryClass = "astro-ph.CO",
    doi = "10.1038/s41586-018-0606-0",
    journal = "Nature",
    volume = "562",
    number = "7728",
    pages = "545--547",
    year = "2018"
}

@article{LIGOScientific:2018gmd,
    author = "{Fishbach}, M. and {Gray}, R. and {Maga{\~n}a Hernandez}, I. and others",
    collaboration = "LIGO Scientific, Virgo",
    title = "{A Standard Siren Measurement of the Hubble Constant from GW170817 without the Electromagnetic Counterpart}",
    eprint = "1807.05667",
    archivePrefix = "arXiv",
    primaryClass = "astro-ph.CO",
    reportNumber = "LIGO-P1800192",
    doi = "10.3847/2041-8213/aaf96e",
    journal = "ApJL",
    volume = "871",
    number = "1",
    pages = "L13",
    year = "2019"
}

@article{Gray:2019ksv,
    author = "{Gray}, Rachel and {Hernandez}, Ignacio Maga{\~n}a and {Qi}, Hong and others",
    title = "{Cosmological inference using gravitational wave standard sirens: A mock data analysis}",
    eprint = "1908.06050",
    archivePrefix = "arXiv",
    primaryClass = "gr-qc",
    reportNumber = "LIGO-P1900017",
    doi = "10.1103/PhysRevD.101.122001",
    journal = "Phys. Rev. D",
    volume = "101",
    number = "12",
    pages = "122001",
    year = "2020"
}

@article{Finke:2021aom,
    author = "Finke, Andreas and Foffa, Stefano and Iacovelli, Francesco and Maggiore, Michele and Mancarella, Michele",
    title = "{Cosmology with LIGO/Virgo dark sirens: Hubble parameter and modified gravitational wave propagation}",
    eprint = "2101.12660",
    archivePrefix = "arXiv",
    primaryClass = "astro-ph.CO",
    doi = "10.1088/1475-7516/2021/08/026",
    journal = "JCAP",
    volume = "08",
    pages = "026",
    year = "2021"
}

@article{Gray:2021sew,
    author = "Gray, Rachel and Messenger, Chris and Veitch, John",
    title = "{A pixelated approach to galaxy catalogue incompleteness: improving the dark siren measurement of the Hubble constant}",
    eprint = "2111.04629",
    archivePrefix = "arXiv",
    primaryClass = "astro-ph.CO",
    doi = "10.1093/mnras/stac366",
    journal = "MNRAS",
    volume = "512",
    number = "1",
    pages = "1127--1140",
    year = "2022"
}

@article{Belfiglio:2022cnd,
    author = "Belfiglio, Alessio and Luongo, Orlando and Mancini, Stefano",
    title = "{Geometric corrections to cosmological entanglement}",
    eprint = "2201.12299",
    archivePrefix = "arXiv",
    primaryClass = "gr-qc",
    doi = "10.1103/PhysRevD.105.123523",
    journal = "Phys. Rev. D",
    volume = "105",
    number = "12",
    pages = "123523",
    year = "2022"
}

@article{Gair:2022zsa,
    author = "Gair, Jonathan R. and Ghosh, Archisman and Gray, Rachel and others",
    title = "{The Hitchhiker{\textquoteright}s Guide to the Galaxy Catalog Approach for Dark Siren Gravitational-wave Cosmology}",
    eprint = "2212.08694",
    archivePrefix = "arXiv",
    primaryClass = "gr-qc",
    doi = "10.3847/1538-3881/acca78",
    journal = "AJ",
    volume = "166",
    number = "1",
    pages = "22",
    year = "2023"
}

@article{Taylor:2011fs,
    author = "Taylor, Stephen R. and Gair, Jonathan R. and Mandel, Ilya",
    title = "{Hubble without the Hubble: Cosmology using advanced gravitational-wave detectors alone}",
    eprint = "1108.5161",
    archivePrefix = "arXiv",
    primaryClass = "gr-qc",
    doi = "10.1103/PhysRevD.85.023535",
    journal = "Phys. Rev. D",
    volume = "85",
    pages = "023535",
    year = "2012"
}

@article{Farr:2019rap,
    author = "Farr, Will M.",
    title = "{Accuracy Requirements for Empirically-Measured Selection Functions}",
    eprint = "1904.10879",
    archivePrefix = "arXiv",
    primaryClass = "astro-ph.IM",
    doi = "10.3847/2515-5172/ab1d5f",
    journal = "Research Notes of the AAS",
    volume = "3",
    number = "5",
    pages = "66",
    year = "2019"
}

@article{Mastrogiovanni:2021wsd,
    author = "Mastrogiovanni, S. and Leyde, K. and Karathanasis, C. and Chassande-Mottin, E. and Steer, D. A. and Gair, J. and Ghosh, A. and Gray, R. and Mukherjee, S. and Rinaldi, S.",
    title = "{On the importance of source population models for gravitational-wave cosmology}",
    eprint = "2103.14663",
    archivePrefix = "arXiv",
    primaryClass = "gr-qc",
    doi = "10.1103/PhysRevD.104.062009",
    journal = "Phys. Rev. D",
    volume = "104",
    number = "6",
    pages = "062009",
    year = "2021"
}

@article{Mastrogiovanni:2023emh,
    author = "Mastrogiovanni, Simone and Laghi, Danny and Gray, Rachel and Santoro, Giada Caneva and Ghosh, Archisman and Karathanasis, Christos and Leyde, Konstantin and Steer, Daniele A. and Perries, Stephane and Pierra, Gregoire",
    title = "{Joint population and cosmological properties inference with gravitational waves standard sirens and galaxy surveys}",
    eprint = "2305.10488",
    archivePrefix = "arXiv",
    primaryClass = "astro-ph.CO",
    doi = "10.1103/PhysRevD.108.042002",
    journal = "Phys. Rev. D",
    volume = "108",
    number = "4",
    pages = "042002",
    year = "2023"
}

@article{Mastrogiovanni:2023zbw,
    author = "Mastrogiovanni, Simone and Pierra, Gr{\'e}goire and Perri{\`e}s, St{\'e}phane and Laghi, Danny and Caneva Santoro, Giada and Ghosh, Archisman and Gray, Rachel and Karathanasis, Christos and Leyde, Konstantin",
    title = "{ICAROGW: A python package for inference of astrophysical population properties of noisy, heterogeneous, and incomplete observations}",
    eprint = "2305.17973",
    archivePrefix = "arXiv",
    primaryClass = "astro-ph.CO",
    doi = "10.1051/0004-6361/202347007",
    journal = "A\&A",
    volume = "682",
    pages = "A167",
    year = "2024"
}

@article{Gray:2023wgj,
    author = "Gray, Rachel and Beirnaert, Freija and Karathanasis, Christos and others",
    title = "{Joint cosmological and gravitational-wave population inference using dark sirens and galaxy catalogues}",
    eprint = "2308.02281",
    archivePrefix = "arXiv",
    primaryClass = "astro-ph.CO",
    doi = "10.1088/1475-7516/2023/12/023",
    journal = "JCAP",
    volume = "12",
    pages = "023",
    year = "2023"
}

@article{Borghi:2023opd,
    author = "Borghi, Nicola and Mancarella, Michele and Moresco, Michele and Tagliazucchi, Matteo and Iacovelli, Francesco and Cimatti, Andrea and Maggiore, Michele",
    title = "{Cosmology and Astrophysics with Standard Sirens and Galaxy Catalogs in View of Future Gravitational Wave Observations}",
    eprint = "2312.05302",
    archivePrefix = "arXiv",
    primaryClass = "astro-ph.CO",
    doi = "10.3847/1538-4357/ad20eb",
    journal = "ApJ",
    volume = "964",
    number = "2",
    pages = "191",
    year = "2024"
}

@article{Tagliazucchi:2025ofb,
    author = "Tagliazucchi, Matteo and Moresco, Michele and Borghi, Nicola and Fiebig, Manfred",
    title = "{Accelerating the standard siren method: Improved constraints on modified gravitational-wave propagation with future data}",
    eprint = "2504.02034",
    archivePrefix = "arXiv",
    primaryClass = "astro-ph.CO",
    doi = "10.1051/0004-6361/202554827",
    journal = "A\&A",
    volume = "702",
    pages = "A244",
    year = "2025"
}

@article{Mancarella:2025uat,
    author = "Mancarella, Michele and Gerosa, Davide",
    title = "{Sampling the full hierarchical population posterior distribution in gravitational-wave astronomy}",
    eprint = "2502.12156",
    archivePrefix = "arXiv",
    primaryClass = "gr-qc",
    doi = "10.1103/PhysRevD.111.103012",
    journal = "Phys. Rev. D",
    volume = "111",
    number = "10",
    pages = "103012",
    year = "2025"
}

@article{LIGOScientific:2021aug,
    author = "Abbott, R. and Abe, H. and Acernese, F. and others",
    collaboration = "LIGO Scientific, Virgo, KAGRA",
    title = "{Constraints on the Cosmic Expansion History from GWTC{\textendash}3}",
    eprint = "2111.03604",
    archivePrefix = "arXiv",
    primaryClass = "astro-ph.CO",
    reportNumber = "LIGO-P2100185-v6, LIGO-P2100185-v5",
    doi = "10.3847/1538-4357/ac74bb",
    journal = "ApJ",
    volume = "949",
    number = "2",
    pages = "76",
    year = "2023"
}

@article{LIGOScientific:2019zcs,
    author = "{Abbott}, B.~P. and {Abbott}, R. and {Abbott}, T.~D. and others",
    collaboration = "LIGO Scientific, Virgo, VIRGO",
    title = "{A Gravitational-wave Measurement of the Hubble Constant Following the Second Observing Run of Advanced LIGO and Virgo}",
    eprint = "1908.06060",
    archivePrefix = "arXiv",
    primaryClass = "astro-ph.CO",
    reportNumber = "LIGO-P1900015",
    doi = "10.3847/1538-4357/abdcb7",
    journal = "ApJ",
    volume = "909",
    number = "2",
    pages = "218",
    year = "2021"
}

@article{LIGOScientific:2025jau,
    author = "Abac, A. G. and Abouelfettouh, I. and Acernese, F. and others",
    collaboration = "LIGO Scientific, VIRGO, KAGRA",
    title = "{GWTC-4.0: Constraints on the Cosmic Expansion Rate and Modified Gravitational-wave Propagation}",
    eprint = "2509.04348",
    note = "{ApJL}, submitted",
    archivePrefix = "arXiv",
    primaryClass = "astro-ph.CO",
    reportNumber = "LIGO-P2400152",
    month = "9",
    year = "2025"
}

@article{LIGOScientific:2018mvr,
    author = "Abbott, B. P. and Abbott, R. and Abbott, T. D. and others",
    collaboration = "LIGO Scientific, Virgo",
    title = "{GWTC-1: A Gravitational-Wave Transient Catalog of Compact Binary Mergers Observed by LIGO and Virgo during the First and Second Observing Runs}",
    eprint = "1811.12907",
    archivePrefix = "arXiv",
    primaryClass = "astro-ph.HE",
    reportNumber = "LIGO-P1800307",
    doi = "10.1103/PhysRevX.9.031040",
    journal = "Phys. Rev. X",
    volume = "9",
    number = "3",
    pages = "031040",
    year = "2019"
}

@article{LIGOScientific:2020ibl,
    author = "{Abbott}, R. and {Abbott}, T.~D. and {Abraham}, S. and others",
    collaboration = "LIGO Scientific, Virgo",
    title = "{GWTC-2: Compact Binary Coalescences Observed by LIGO and Virgo During the First Half of the Third Observing Run}",
    eprint = "2010.14527",
    archivePrefix = "arXiv",
    primaryClass = "gr-qc",
    reportNumber = "P2000061",
    doi = "10.1103/PhysRevX.11.021053",
    journal = "Phys. Rev. X",
    volume = "11",
    pages = "021053",
    year = "2021"
}

@article{KAGRA:2021vkt,
    author = "Abbott, R. and Abbott, T. D. and Acernese, F. and others",
    collaboration = "KAGRA, VIRGO, LIGO Scientific",
    title = "{GWTC-3: Compact Binary Coalescences Observed by LIGO and Virgo during the Second Part of the Third Observing Run}",
    eprint = "2111.03606",
    archivePrefix = "arXiv",
    primaryClass = "gr-qc",
    reportNumber = "LIGO-P2000318",
    doi = "10.1103/PhysRevX.13.041039",
    journal = "Phys. Rev. X",
    volume = "13",
    number = "4",
    pages = "041039",
    year = "2023"
}

@article{LIGOScientific:2025slb,
    author = "Abac, A. G. and Abouelfettouh, I. and Acernese, F. and others",
    collaboration = "LIGO Scientific, VIRGO, KAGRA",
    title = "{GWTC-4.0: Updating the Gravitational-Wave Transient Catalog with Observations from the First Part of the Fourth LIGO-Virgo-KAGRA Observing Run}",
    eprint = "2508.18082",
    note = "{ApJL}, submitted",
    archivePrefix = "arXiv",
    primaryClass = "gr-qc",
    reportNumber = "LIGO-P2400386",
    month = "8",
    year = "2025"
}

@article{Hild:2010id,
    author = "{Hild}, S. and {Abernathy}, M. and {Acernese}, F. and others",
    title = "{Sensitivity Studies for Third-Generation Gravitational Wave Observatories}",
    eprint = "1012.0908",
    archivePrefix = "arXiv",
    primaryClass = "gr-qc",
    doi = "10.1088/0264-9381/28/9/094013",
    journal = "Class. Quant. Grav.",
    volume = "28",
    pages = "094013",
    year = "2011"
}

@article{Punturo:2010zz,
    author = "Punturo, M. and Abernathy, M. and Acernese, F. and others",
    editor = "Ricci, Fulvio",
    title = "{The Einstein Telescope: A third-generation gravitational wave observatory}",
    doi = "10.1088/0264-9381/27/19/194002",
    journal = "Class. Quant. Grav.",
    volume = "27",
    pages = "194002",
    year = "2010"
}

@article{ET:2019dnz,
    author = "{Maggiore}, Michele and {Van Den Broeck}, Chris and {Bartolo}, Nicola and others",
    collaboration = "ET",
    title = "{Science Case for the Einstein Telescope}",
    eprint = "1912.02622",
    archivePrefix = "arXiv",
    primaryClass = "astro-ph.CO",
    doi = "10.1088/1475-7516/2020/03/050",
    journal = "JCAP",
    volume = "03",
    pages = "050",
    year = "2020"
}

@article{Reitze:2019iox,
    author = "{Reitze}, David and {Adhikari}, Rana X. and {Ballmer}, Stefan and others",
    title = "{Cosmic Explorer: The U.S. Contribution to Gravitational-Wave Astronomy beyond LIGO}",
    eprint = "1907.04833",
    archivePrefix = "arXiv",
    primaryClass = "astro-ph.IM",
    reportNumber = "LIGO-P1900316",
    journal = "Bull. Am. Astron. Soc.",
    volume = "51",
    number = "7",
    pages = "035",
    year = "2019"
}

@article{Evans:2021gyd,
    author = "{Evans}, Matthew and {Adhikari}, Rana X and {Afle}, Chaitanya and others",
    title = "{A Horizon Study for Cosmic Explorer: Science, Observatories, and Community}",
    eprint = "2109.09882",
    archivePrefix = "arXiv",
    primaryClass = "astro-ph.IM",
    reportNumber = "CE-P2100003-v7, Cosmic Explorer technical report CE-P2100003-v6",
    month = "9",
    year = "2021"
}

@article{Iacovelli:2022bbs,
    author = "Iacovelli, Francesco and Mancarella, Michele and Foffa, Stefano and Maggiore, Michele",
    title = "{Forecasting the Detection Capabilities of Third-generation Gravitational-wave Detectors Using GWFAST}",
    eprint = "2207.02771",
    archivePrefix = "arXiv",
    primaryClass = "gr-qc",
    doi = "10.3847/1538-4357/ac9cd4",
    journal = "ApJ",
    volume = "941",
    number = "2",
    pages = "208",
    year = "2022"
}

@article{Branchesi:2023mws,
    author = "{Branchesi}, Marica and {Maggiore}, Michele and {Alonso}, David and others",
    title = "{Science with the Einstein Telescope: a comparison of different designs}",
    eprint = "2303.15923",
    archivePrefix = "arXiv",
    primaryClass = "gr-qc",
    reportNumber = "ET-0084A-23",
    doi = "10.1088/1475-7516/2023/07/068",
    journal = "JCAP",
    volume = "07",
    pages = "068",
    year = "2023"
}

@article{ET:2025xjr,
    author = "{Abac}, Adrian and {Abramo}, Raul and {Albanesi}, Simone and others",
    collaboration = "ET",
    title = "{The Science of the Einstein Telescope}",
    eprint = "2503.12263",
    archivePrefix = "arXiv",
    primaryClass = "gr-qc",
    reportNumber = "ET-0036C-25",
    doi = "10.1088/1475-7516/2026/03/081",
    journal = "JCAP",
    volume = "03",
    pages = "081",
    year = "2026"
}

@article{Mandel:2018mve,
    author = "Mandel, Ilya and Farr, Will M. and Gair, Jonathan R.",
    title = "{Extracting distribution parameters from multiple uncertain observations with selection biases}",
    eprint = "1809.02063",
    archivePrefix = "arXiv",
    primaryClass = "physics.data-an",
    doi = "10.1093/mnras/stz896",
    journal = "MNRAS",
    volume = "486",
    number = "1",
    pages = "1086--1093",
    year = "2019"
}

@incollection{Vitale:2020aaz,
    author="Vitale, Salvatore and Gerosa, Davide and Farr, Will M. and Taylor, Stephen R.", 
    editor="Bambi, Cosimo and Katsanevas, Stavros and Kokkotas, Konstantinos D.",
    title="Inferring the Properties of a Population of Compact Binaries in Presence of Selection Effects",
    booktitle="Handbook of Gravitational Wave Astronomy",
    year="2020",
    publisher="Springer Singapore",
    address="Singapore",
    pages="1--60",
    isbn="978-981-15-4702-7",
    doi="10.1007/978-981-15-4702-7_45-1",
}

@article{Pierra:2025hoc,
    author = "Pierra, Gr{\'e}goire and Colombo, Alberto and Mastrogiovanni, Simone",
    title = "{Non-Parametric Reconstruction of the Hubble Parameter from the Fourth Gravitational Wave Transient Catalog and DESI Baryonic Acoustic Oscillations}",
    eprint = "2511.11795",
    note = {{CQG, accepted}},
    archivePrefix = "arXiv",
    primaryClass = "astro-ph.CO",
    month = "11",
    year = "2025"
}

@article{Mancarella:2021ecn,
    author = "Mancarella, Michele and Genoud-Prachex, Edwin and Maggiore, Michele",
    title = "{Cosmology and modified gravitational wave propagation from binary black hole population models}",
    eprint = "2112.05728",
    archivePrefix = "arXiv",
    primaryClass = "gr-qc",
    doi = "10.1103/PhysRevD.105.064030",
    journal = "Phys. Rev. D",
    volume = "105",
    number = "6",
    pages = "064030",
    year = "2022"
}

@article{Tiwari:2017ndi,
    author = "Tiwari, Vaibhav",
    title = "{Estimation of the Sensitive Volume for Gravitational-wave Source Populations Using Weighted Monte Carlo Integration}",
    eprint = "1712.00482",
    archivePrefix = "arXiv",
    primaryClass = "astro-ph.HE",
    doi = "10.1088/1361-6382/aac89d",
    journal = "Class. Quant. Grav.",
    volume = "35",
    number = "14",
    pages = "145009",
    year = "2018"
}

@article{LIGOScientific:2020kqk,
    author = "{Abbott}, R. and {Abbott}, T.~D. and {Abraham}, S.  others",
    collaboration = "LIGO Scientific, Virgo",
    title = "{Population Properties of Compact Objects from the Second LIGO-Virgo Gravitational-Wave Transient Catalog}",
    eprint = "2010.14533",
    archivePrefix = "arXiv",
    primaryClass = "astro-ph.HE",
    reportNumber = "LIGO-P2000077",
    doi = "10.3847/2041-8213/abe949",
    journal = "ApJL",
    volume = "913",
    number = "1",
    pages = "L7",
    year = "2021"
}

@article{Madau:2014bja,
    author = "Madau, Piero and Dickinson, Mark",
    title = "{Cosmic Star Formation History}",
    eprint = "1403.0007",
    archivePrefix = "arXiv",
    primaryClass = "astro-ph.CO",
    doi = "10.1146/annurev-astro-081811-125615",
    journal = "Ann. Rev. A\&A",
    volume = "52",
    pages = "415--486",
    year = "2014"
}

@article{London:2017bcn,
    author = "London, Lionel and Khan, Sebastian and Fauchon-Jones, Edward and Garc{\'\i}a, Cecilio and Hannam, Mark and Husa, Sascha and Jim{\'e}nez-Forteza, Xisco and Kalaghatgi, Chinmay and Ohme, Frank and Pannarale, Francesco",
    title = "{First higher-multipole model of gravitational waves from spinning and coalescing black-hole binaries}",
    eprint = "1708.00404",
    archivePrefix = "arXiv",
    primaryClass = "gr-qc",
    doi = "10.1103/PhysRevLett.120.161102",
    journal = "Phys. Rev. Lett.",
    volume = "120",
    number = "16",
    pages = "161102",
    year = "2018"
}

@article{Iacovelli:2022mbg,
    author = "Iacovelli, Francesco and Mancarella, Michele and Foffa, Stefano and Maggiore, Michele",
    title = "{GWFAST: A Fisher Information Matrix Python Code for Third-generation Gravitational-wave Detectors}",
    eprint = "2207.06910",
    archivePrefix = "arXiv",
    primaryClass = "astro-ph.IM",
    doi = "10.3847/1538-4365/ac9129",
    journal = "ApJ Supp.",
    volume = "263",
    number = "1",
    pages = "2",
    year = "2022"
}

@article{Foreman-Mackey:2012any,
    author = "Foreman-Mackey, Daniel and Hogg, David W. and Lang, Dustin and Goodman, Jonathan",
    title = "{emcee: The MCMC Hammer}",
    eprint = "1202.3665",
    archivePrefix = "arXiv",
    primaryClass = "astro-ph.IM",
    doi = "10.1086/670067",
    journal = "Publ. Astron. Soc. Pac.",
    volume = "125",
    pages = "306--312",
    year = "2013"
}

@article{Talbot:2023pex,
    author = "Talbot, Colm and Golomb, Jacob",
    title = "{Growing pains: understanding the impact of likelihood uncertainty on hierarchical Bayesian inference for gravitational-wave astronomy}",
    eprint = "2304.06138",
    archivePrefix = "arXiv",
    primaryClass = "astro-ph.IM",
    doi = "10.1093/mnras/stad2968",
    journal = "MNRAS",
    volume = "526",
    number = "3",
    pages = "3495--3503",
    year = "2023"
}

@article{Heinzel:2025ogf,
    author = "Heinzel, Jack and Vitale, Salvatore",
    title = "{When (not) to trust Monte Carlo approximations for hierarchical Bayesian inference}",
    eprint = "2509.07221",
    archivePrefix = "arXiv",
    primaryClass = "astro-ph.HE",
    month = "9",
    year = "2025"
}

@article{Karamanis:2022alw,
    author = "Karamanis, Minas and Beutler, Florian and Peacock, John A. and Nabergoj, David and Seljak, Uros",
    title = "{Accelerating astronomical and cosmological inference with preconditioned Monte Carlo}",
    eprint = "2207.05652",
    archivePrefix = "arXiv",
    primaryClass = "astro-ph.IM",
    doi = "10.1093/mnras/stac2272",
    journal = "MNRAS",
    volume = "516",
    number = "2",
    pages = "1644--1653",
    year = "2022"
}

@article{Karamanis:2022ksp,
    author = "Karamanis, Minas and Nabergoj, David and Beutler, Florian and Peacock, John A. and Seljak, Uros",
    title = "{pocoMC: A Python package for accelerated Bayesian inference in astronomy and cosmology}",
    eprint = "2207.05660",
    archivePrefix = "arXiv",
    primaryClass = "astro-ph.IM",
    doi = "10.21105/joss.04634",
    journal = "J. Open Source Softw.",
    volume = "7",
    number = "79",
    pages = "4634",
    year = "2022"
}

@article{Ye:2021klk,
    author = "Ye, Christine and Fishbach, Maya",
    title = "{Cosmology with standard sirens at cosmic noon}",
    eprint = "2103.14038",
    archivePrefix = "arXiv",
    primaryClass = "astro-ph.CO",
    doi = "10.1103/PhysRevD.104.043507",
    journal = "Phys. Rev. D",
    volume = "104",
    number = "4",
    pages = "043507",
    year = "2021"
}

@article{Mukherjee:2021rtw,
    author = "Mukherjee, Suvodip",
    title = "{The redshift dependence of black hole mass distribution: is it reliable for standard sirens cosmology?}",
    eprint = "2112.10256",
    archivePrefix = "arXiv",
    primaryClass = "astro-ph.CO",
    doi = "10.1093/mnras/stac2152",
    journal = "MNRAS",
    volume = "515",
    number = "4",
    pages = "5495--5505",
    year = "2022"
}

@article{Karathanasis:2022rtr,
    author = "Karathanasis, Christos and Mukherjee, Suvodip and Mastrogiovanni, Simone",
    title = "{Binary black holes population and cosmology in new lights: signature of PISN mass and formation channel in GWTC-3}",
    eprint = "2204.13495",
    archivePrefix = "arXiv",
    primaryClass = "astro-ph.CO",
    doi = "10.1093/mnras/stad1373",
    journal = "MNRAS",
    volume = "523",
    number = "3",
    pages = "4539--4555",
    year = "2023"
}

@article{Pierra:2023deu,
    author = "Pierra, Gr{\'e}goire and Mastrogiovanni, Simone and Perri{\`e}s, St{\'e}phane and Mapelli, Michela",
    title = "{Study of systematics on the cosmological inference of the Hubble constant from gravitational wave standard sirens}",
    eprint = "2312.11627",
    archivePrefix = "arXiv",
    primaryClass = "astro-ph.CO",
    reportNumber = "LIGO-P2300442",
    doi = "10.1103/PhysRevD.109.083504",
    journal = "Phys. Rev. D",
    volume = "109",
    number = "8",
    pages = "083504",
    year = "2024"
}

@article{Agarwal:2024hld,
    author = "Agarwal, Aman and others",
    title = "{Blinded Mock Data Challenge for Gravitational-wave Cosmology. I. Assessing the Robustness of Methods Using Binary Black Hole Mass Spectrum}",
    eprint = "2412.14244",
    archivePrefix = "arXiv",
    primaryClass = "astro-ph.CO",
    doi = "10.3847/1538-4357/adda3a",
    journal = "ApJ",
    volume = "987",
    number = "1",
    pages = "47",
    year = "2025"
}

@article{Wong:2020ise,
    author = "Wong, Kaze W. K. and Breivik, Katelyn and Kremer, Kyle and Callister, Thomas",
    title = "{Joint constraints on the field-cluster mixing fraction, common envelope efficiency, and globular cluster radii from a population of binary hole mergers via deep learning}",
    eprint = "2011.03564",
    archivePrefix = "arXiv",
    primaryClass = "astro-ph.HE",
    doi = "10.1103/PhysRevD.103.083021",
    journal = "Phys. Rev. D",
    volume = "103",
    number = "8",
    pages = "083021",
    year = "2021"
}

@article{Zevin:2022bfa,
    author = "Zevin, Michael and Holz, Daniel E.",
    title = "{Avoiding a Cluster Catastrophe: Retention Efficiency and the Binary Black Hole Mass Spectrum}",
    eprint = "2205.08549",
    archivePrefix = "arXiv",
    primaryClass = "astro-ph.HE",
    doi = "10.3847/2041-8213/ac853d",
    journal = "ApJL",
    volume = "935",
    pages = "L20",
    year = "2022"
}

@article{Li:2023yyt,
    author = "Li, Yin-Jie and Wang, Yuan-Zhu and Tang, Shao-Peng and Fan, Yi-Zhong",
    title = "{Resolving the Stellar-Collapse and Hierarchical-Merger Origins of the Coalescing Black Holes}",
    eprint = "2303.02973",
    archivePrefix = "arXiv",
    primaryClass = "astro-ph.HE",
    doi = "10.1103/PhysRevLett.133.051401",
    journal = "Phys. Rev. Lett.",
    volume = "133",
    number = "5",
    pages = "051401",
    year = "2024"
}

@article{Torniamenti:2024uxl,
    author = "Torniamenti, Stefano and Mapelli, Michela and P{\'e}rigois, Carole and Sedda, Manuel Arca and Artale, Maria Celeste and Dall'Amico, Marco and Vaccaro, Maria Paola",
    title = "{Hierarchical binary black hole mergers in globular clusters: Mass function and evolution with redshift}",
    eprint = "2401.14837",
    archivePrefix = "arXiv",
    primaryClass = "astro-ph.HE",
    doi = "10.1051/0004-6361/202449272",
    journal = "A\&A",
    volume = "688",
    pages = "A148",
    year = "2024"
}

@article{Fishbach:2024hfw,
    author = "Fishbach, Maya",
    title = "{Probing cosmic chemical enrichment with next-generation gravitational-wave observatories}",
    eprint = "2411.08658",
    archivePrefix = "arXiv",
    primaryClass = "astro-ph.HE",
    doi = "10.1088/1361-6382/adaf70",
    journal = "Class. Quant. Grav.",
    volume = "42",
    number = "5",
    pages = "055009",
    year = "2025"
}

@article{Li:2024rmi,
    author = "Li, Yin-Jie and Tang, Shao-Peng and Wang, Yuan-Zhu and Fan, Yi-Zhong",
    title = "{Multispectral Sirens: Gravitational-wave Cosmology with (Multi-) Subpopulations of Binary Black Holes}",
    eprint = "2406.11607",
    archivePrefix = "arXiv",
    primaryClass = "astro-ph.CO",
    doi = "10.3847/1538-4357/ad888b",
    journal = "ApJ",
    volume = "976",
    number = "2",
    pages = "153",
    year = "2024"
}

@article{Woosley:2021xba,
    author = "Woosley, S. E. and Heger, Alexander",
    title = "{The Pair-Instability Mass Gap for Black Holes}",
    eprint = "2103.07933",
    archivePrefix = "arXiv",
    primaryClass = "astro-ph.SR",
    doi = "10.3847/2041-8213/abf2c4",
    journal = "ApJL",
    volume = "912",
    number = "2",
    pages = "L31",
    year = "2021"
}

@article{vanSon:2021zpk,
    author = "van Son, L. A. C. and de Mink, S. E. and Callister, T. and Justham, S. and Renzo, M. and Wagg, T. and Broekgaarden, F. S. and Kummer, F. and Pakmor, R. and Mandel, I.",
    title = "{The Redshift Evolution of the Binary Black Hole Merger Rate: A Weighty Matter}",
    eprint = "2110.01634",
    archivePrefix = "arXiv",
    primaryClass = "astro-ph.HE",
    doi = "10.3847/1538-4357/ac64a3",
    journal = "ApJ",
    volume = "931",
    number = "1",
    pages = "17",
    year = "2022"
}

@article{Antonini:2016gqe,
    author = "Antonini, Fabio and Rasio, Frederic A.",
    title = "{Merging black hole binaries in galactic nuclei: implications for advanced-LIGO detections}",
    eprint = "1606.04889",
    archivePrefix = "arXiv",
    primaryClass = "astro-ph.HE",
    doi = "10.3847/0004-637X/831/2/187",
    journal = "ApJ",
    volume = "831",
    number = "2",
    pages = "187",
    year = "2016"
}

@article{Mapelli:2021gyv,
    author = "Mapelli, Michela and Bouffanais, Yann and Santoliquido, Filippo and Sedda, Manuel Arca and Artale, M. Celeste",
    title = "{The cosmic evolution of binary black holes in young, globular, and nuclear star clusters: rates, masses, spins, and mixing fractions}",
    eprint = "2109.06222",
    archivePrefix = "arXiv",
    primaryClass = "astro-ph.HE",
    doi = "10.1093/mnras/stac422",
    journal = "MNRAS",
    volume = "511",
    number = "4",
    pages = "5797--5816",
    year = "2022"
}

@article{Gerosa:2021mno,
    author = "Gerosa, Davide and Fishbach, Maya",
    title = "{Hierarchical mergers of stellar-mass black holes and their gravitational-wave signatures}",
    eprint = "2105.03439",
    archivePrefix = "arXiv",
    primaryClass = "astro-ph.HE",
    doi = "10.1038/s41550-021-01398-w",
    journal = "Nature Astron.",
    volume = "5",
    number = "8",
    pages = "749--760",
    year = "2021"
}

@article{Ye:2024ypm,
    author = "Ye, Claire S. and Fishbach, Maya",
    title = "{The Redshift Evolution of the Binary Black Hole Mass Distribution from Dense Star Clusters}",
    eprint = "2402.12444",
    archivePrefix = "arXiv",
    primaryClass = "astro-ph.HE",
    doi = "10.3847/1538-4357/ad3ba8",
    journal = "ApJ",
    volume = "967",
    number = "1",
    pages = "62",
    year = "2024"
}

@article{Ford:2021kcw,
    author = "Ford, K. E. Saavik and McKernan, Barry",
    title = "{Binary black hole merger rates in AGN discs versus nuclear star clusters: loud beats quiet}",
    eprint = "2109.03212",
    archivePrefix = "arXiv",
    primaryClass = "astro-ph.HE",
    doi = "10.1093/mnras/stac2861",
    journal = "MNRAS",
    volume = "517",
    number = "4",
    pages = "5827--5834",
    year = "2022"
}

@article{Rinaldi:2023bbd,
    author = "Rinaldi, Stefano and Del Pozzo, Walter and Mapelli, Michela and Lorenzo-Medina, Ana and Dent, Thomas",
    title = "{Evidence of evolution of the black hole mass function with redshift}",
    eprint = "2310.03074",
    archivePrefix = "arXiv",
    primaryClass = "astro-ph.HE",
    doi = "10.1051/0004-6361/202348161",
    journal = "A\&A",
    volume = "684",
    pages = "A204",
    year = "2024"
}

@article{Heinzel:2024hva,
    author = "Heinzel, Jack and Mould, Matthew and Vitale, Salvatore",
    title = "{Nonparametric analysis of correlations in the binary black hole population with LIGO-Virgo-KAGRA data}",
    eprint = "2406.16844",
    archivePrefix = "arXiv",
    primaryClass = "astro-ph.HE",
    doi = "10.1103/PhysRevD.111.L061305",
    journal = "Phys. Rev. D",
    volume = "111",
    number = "6",
    pages = "L061305",
    year = "2025"
}

@article{Farah:2024xub,
    author = "Farah, Amanda M. and Callister, Thomas A. and Ezquiaga, Jose Mar{\'\i}a and Zevin, Michael and Holz, Daniel E.",
    title = "{No Need to Know: Toward Astrophysics-free Gravitational-wave Cosmology}",
    eprint = "2404.02210",
    archivePrefix = "arXiv",
    primaryClass = "astro-ph.CO",
    doi = "10.3847/1538-4357/ad9253",
    journal = "ApJ",
    volume = "978",
    number = "2",
    pages = "153",
    year = "2025"
}

@article{Sadiq:2025aog,
    author = "Sadiq, Jam and Dent, Thomas and Lorenzo-Medina, Ana",
    title = "{Probing evolution in the black hole spectrum with gravitational waves catalogs}",
    eprint = "2502.06451",
    archivePrefix = "arXiv",
    primaryClass = "gr-qc",
    doi = "10.1103/j9fq-nkl3",
    journal = "Phys. Rev. D",
    volume = "112",
    number = "8",
    pages = "083028",
    year = "2025"
}

@article{Farr:2019twy,
    author = "Farr, Will M. and Fishbach, Maya and Ye, Jiani and Holz, Daniel",
    title = "{A Future Percent-Level Measurement of the Hubble Expansion at Redshift 0.8 With Advanced LIGO}",
    eprint = "1908.09084",
    archivePrefix = "arXiv",
    primaryClass = "astro-ph.CO",
    reportNumber = "LIGO P1900252",
    doi = "10.3847/2041-8213/ab4284",
    journal = "ApJL",
    volume = "883",
    number = "2",
    pages = "L42",
    year = "2019"
}

@article{Moresco:2016mzx,
    author = "Moresco, Michele and Pozzetti, Lucia and Cimatti, Andrea and Jimenez, Raul and Maraston, Claudia and Verde, Licia and Thomas, Daniel and Citro, Annalisa and Tojeiro, Rita and Wilkinson, David",
    title = "{A 6{\%} measurement of the Hubble parameter at $z\sim0.45$: direct evidence of the epoch of cosmic re-acceleration}",
    eprint = "1601.01701",
    archivePrefix = "arXiv",
    primaryClass = "astro-ph.CO",
    doi = "10.1088/1475-7516/2016/05/014",
    journal = "JCAP",
    volume = "05",
    pages = "014",
    year = "2016"
}

@article{Califano:2025qbx,
    author = "Califano, Matteo and De Martino, Ivan and Vernieri, Daniele",
    title = "{Joint estimation of the cosmological model and the mass and redshift distributions of the binary black hole population with the Einstein Telescope}",
    eprint = "2503.19061",
    archivePrefix = "arXiv",
    primaryClass = "astro-ph.CO",
    doi = "10.1103/7nqz-3pzn",
    journal = "Phys. Rev. D",
    volume = "111",
    number = "12",
    pages = "123535",
    year = "2025"
}

@article{Chen:2024gdn,
    author = "Chen, Hsin-Yu and Ezquiaga, Jose Mar{\'\i}a and Gupta, Ish",
    title = "{Cosmography with next-generation gravitational wave detectors}",
    eprint = "2402.03120",
    archivePrefix = "arXiv",
    primaryClass = "gr-qc",
    doi = "10.1088/1361-6382/ad424f",
    journal = "Class. Quant. Grav.",
    volume = "41",
    number = "12",
    pages = "125004",
    year = "2024"
}

@article{Vitale:2016icu,
    author = "Vitale, Salvatore and Evans, Matthew",
    title = "{Parameter estimation for binary black holes with networks of third generation gravitational-wave detectors}",
    eprint = "1610.06917",
    archivePrefix = "arXiv",
    primaryClass = "gr-qc",
    doi = "10.1103/PhysRevD.95.064052",
    journal = "Phys. Rev. D",
    volume = "95",
    number = "6",
    pages = "064052",
    year = "2017"
}

@article{Ezquiaga:2022zkx,
    author = "Ezquiaga, Jose Mar{\'\i}a and Holz, Daniel E.",
    title = "{Spectral Sirens: Cosmology from the Full Mass Distribution of Compact Binaries}",
    eprint = "2202.08240",
    archivePrefix = "arXiv",
    primaryClass = "astro-ph.CO",
    doi = "10.1103/PhysRevLett.129.061102",
    journal = "Phys. Rev. Lett.",
    volume = "129",
    number = "6",
    pages = "061102",
    year = "2022"
}

@article{LIGOScientific:2025pvj,
    author = "Abac, A. G. and Abouelfettouh, I. and Acernese, F. and others",
    collaboration = "LIGO Scientific, VIRGO, KAGRA",
    title = "{GWTC-4.0: Population Properties of Merging Compact Binaries}",
    eprint = "2508.18083",
    archivePrefix = "arXiv",
     note = "{ApJL}, submitted",
    primaryClass = "astro-ph.HE",
    reportNumber = "LIGO-P2400004",
    month = "8",
    year = "2025"
}

@article{Tiwari:2025oah,
    author = "Tiwari, Vaibhav",
    title = "{Population of Binary Black Holes Inferred from One Hundred and Fifty Gravitational Wave Signals}",
    eprint = "2510.25579",
    note = {{MNRAS}, submitted},
    archivePrefix = "arXiv",
    primaryClass = "astro-ph.HE",
    month = "10",
    year = "2025"
}

@article{Pierra:2026ffj,
    author = "Pierra, Gr{\'e}goire and Papadopoulos, Alexander",
    title = "{Heavy Black-Holes Also Matter in Standard Siren Cosmology}",
    eprint = "2601.03257",
    note = {{A\&A, submitted}},
    archivePrefix = "arXiv",
    primaryClass = "astro-ph.CO",
    month = "1",
    year = "2026"
}

@article{Tagliazucchi:2026gxn,
    author = "Tagliazucchi, Matteo and Moresco, Michele and Borghi, Nicola and Ciapetti, Chiara",
    title = "{Mind the peak: Improving cosmological constraints from GWTC-4.0 spectral sirens using semiparametric mass models}",
    eprint = "2601.03347",
    archivePrefix = "arXiv",
    primaryClass = "astro-ph.CO",
    doi = "10.1051/0004-6361/202558756",
    journal = "A\&A",
    volume = "709",
    pages = "A197",
    year = "2026"
}

@article{DESI:2024mwx,
    author = "{Adame}, A.~G. and {Aguilar}, J. and {Ahlen}, S. and others",
    collaboration = "DESI",
    title = "{DESI 2024 VI: cosmological constraints from the measurements of baryon acoustic oscillations}",
    eprint = "2404.03002",
    archivePrefix = "arXiv",
    primaryClass = "astro-ph.CO",
    reportNumber = "FERMILAB-PUB-24-0154-PPD",
    doi = "10.1088/1475-7516/2025/02/021",
    journal = "JCAP",
    volume = "02",
    pages = "021",
    year = "2025"
}

@article{Riess:2021jrx,
    author = "{Riess}, Adam G. and {Yuan}, Wenlong and {Macri}, Lucas M. and others",
    title = "{A Comprehensive Measurement of the Local Value of the Hubble Constant with 1 km s$^{−1}$ Mpc$^{−1}$ Uncertainty from the Hubble Space Telescope and the SH0ES Team}",
    eprint = "2112.04510",
    archivePrefix = "arXiv",
    primaryClass = "astro-ph.CO",
    doi = "10.3847/2041-8213/ac5c5b",
    journal = "ApJL",
    volume = "934",
    number = "1",
    pages = "L7",
    year = "2022"
}

@article{jax2018github,
  author = {James Bradbury and Roy Frostig and Peter Hawkins and Matthew James Johnson and Chris Leary and Dougal Maclaurin and George Necula and Adam Paszke and Jake Vander{P}las and Skye Wanderman-{M}ilne and Qiao Zhang},
  title = {{JAX}: composable transformations of {P}ython+{N}um{P}y programs},
  journal = {\url{http://github.com/jax-ml/jax}},
  version = {0.3.13},
  year = {2018},
}

\begin{appendix}

\FloatBarrier 

\section{Results with \texorpdfstring{$\Omega_{\rm m,0}$}{Om0} fixed}\label{app:Om0fixed-case}

In this Appendix, we present and discuss the results obtained under the assumption that $\Omega_{\rm m,0}$ is held fixed at its fiducial value.

In \Cref{fig:Om0fixed-case_H0}, we compare the constraints derived from the S/N $>$ 75 catalog under two assumptions: one where $\Omega_{\rm m,0}$ is fixed to its fiducial value, and another where it is marginalized over. 
The figure shows only the parameters most correlated with $\Omega_{\rm m,0}$ because the marginalized 1D posteriors for the other parameters remain nearly unchanged.
Fixing $\Omega_{\rm m,0}$ significantly improves the precision of the following parameters:
\begin{itemize}
    \item $H_0$: uncertainty improves from 12\% to 3\%.
    \item $\mu_g$: uncertainty improves from 2.8\% to 1.4\%.
    \item $z_p$: uncertainty improves from 4.5\% to 2.8\%.
    \item $m_{\rm low}$: uncertainty improves from 2.6\% to 1.77\%.
    \item $m_{\rm high}$: uncertainty improves from 2.95\%to 1.85\%.
\end{itemize}

\begin{figure*}[!hb]
    \centering
    \includegraphics[width=0.65\linewidth]{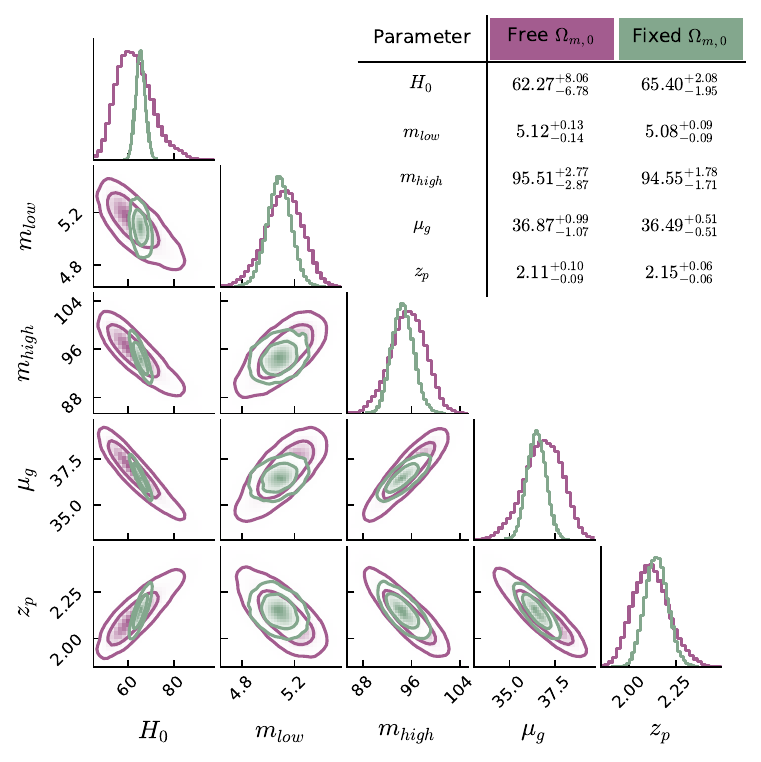}
    \caption{ Comparison of the constraints obtained from the S/N>75 catalog when the matter density parameter $\Omega_{\rm m,0}$ is fixed to its fiducial value (green) and when it is marginalized over (purple). Only parameters significantly correlated with $H_0$ are shown.}
    \label{fig:Om0fixed-case_H0}
\end{figure*}

\newpage
\noindent These improvements are a direct consequence of breaking the degeneracy between $H_0$ and $\Omega_{\rm m,0}$, clearly visible in the joint posteriors shown in \Cref{fig:corner-snr75} and \Cref{fig:results-snr60-vs-snr75} and expected from the luminosity distance-redshift relation.

In \Cref{fig:const_power_Om0fixed} we show  the Pearson correlation coefficients between individual events and the hyperparameters $H_0$, $\mu_g$, and $\gamma$, similarly to the results shown in \Cref{fig:constraining-power}.
We note that:
\begin{itemize}
    \item The first sign reversal, observed in \Cref{fig:constraining-power} in the full parameter analysis at $d_L \sim 6\,\si{\giga\parsec}$ ($z \approx 0.84$), does not occur here. This absence confirms that such reversal is a direct consequence of the anticorrelation between $H_0$ and $\Omega_{\mathrm{m},0}$.
    \item The second sign reversal, observed at $d_L \sim 16\,\si{\giga\parsec}$ ($z \approx 1.83$), is still present even with $\Omega_{\mathrm{m},0}$ fixed.
    This confirms that the mechanism driving this high-distance reversal originates from the changing relationship between distant events and $\gamma$.
\end{itemize}

\begin{figure*}[!hb]
    \centering
    \includegraphics[width=\linewidth]{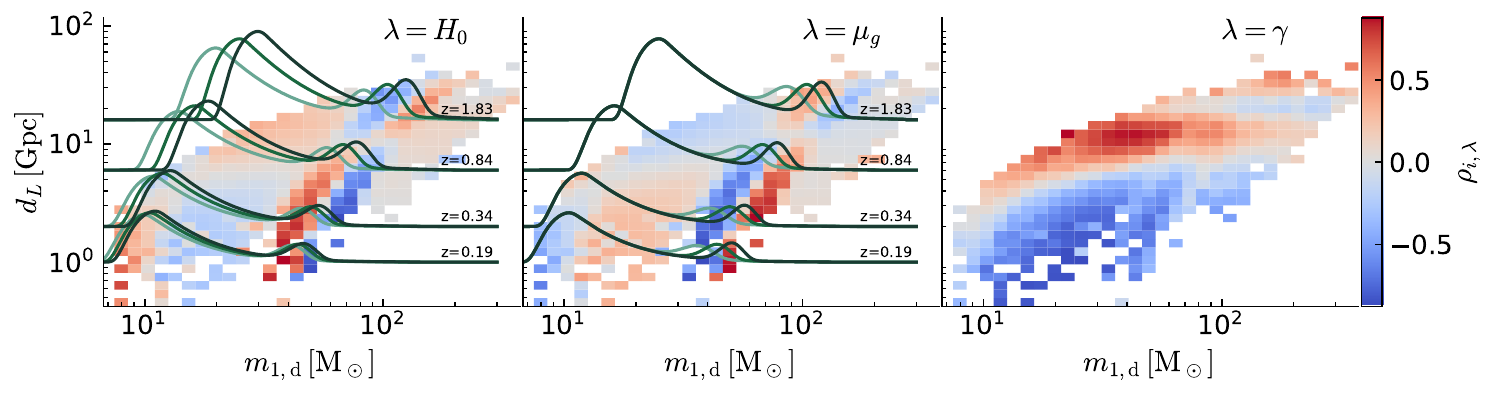}
    \caption{Pearson correlation coefficients between events and hyperparameters, with $\Omega_{\mathrm{m},0}$ fixed. Only the high-distance ($d_L \sim 16\,\si{\giga\parsec}$) sign reversal is present, while the reversal at $d_L \sim 6\,\si{\giga\parsec}$ disappears.}
    \label{fig:const_power_Om0fixed}
\end{figure*}

\end{appendix}
\end{document}